\documentclass[5p,fleqn,times]{elsarticle}

\usepackage{amsmath,amssymb}
\usepackage{mathrsfs}
\usepackage{mathtools}
\usepackage{graphicx}
\usepackage{physics}
\usepackage{subcaption}
\usepackage{pifont}

\newcommand{\Vbar}{\overline{\mathbf{V}}^e}
\newcommand{\Jnorm}[1]{||{#1}||}
\newcommand{\dev}[1]{\mbox{dev}\left({#1}\right)}
\newcommand{\Subref}{\mbox{\scriptsize{ref}}}

\journal{}

\makeatletter
\def\ps@pprintTitle{%
  \let\@oddhead\@empty
  \let\@evenhead\@empty
  \def\@oddfoot{}%
  \let\@evenfoot\@oddfoot}
\makeatletter
\def\ps@pprintTitle{%
  \let\@oddhead\@empty
  \let\@evenhead\@empty
  \def\@oddfoot{}%
  \let\@evenfoot\@oddfoot}
\makeatother
\usepackage{fancyhdr}

\fancypagestyle{pprintTitle}{
\lhead{}\chead{}\rhead{}\lfoot{\textcopyright \small ~British Crown Owned Copyright 2020/AWE}\cfoot{\thepage}\rfoot{}
}

\begin{document}

\begin{frontmatter}

\title{A Diffuse Interface Model of Reactive-fluids and Solid-dynamics}


\author[mymainaddress]{Tim Wallis \corref{mycorrespondingauthor}}
\cortext[mycorrespondingauthor]{Corresponding author}
\ead{tnmw2@cam.ac.uk}

\author[mysecondaryaddress]{Philip T. Barton}
\author[mymainaddress]{Nikolaos Nikiforakis}

\address[mymainaddress]{Department of Physics, University of Cambridge, Cavendish Laboratory, JJ Thomson Avenue, CB3 0HE, UK}
\address[mysecondaryaddress]{AWE Aldermaston, Reading, Berkshire, RG7 4PR, UK}

\begin{abstract}

This article presents a multi-physics methodology for the numerical simulation of physical systems that involve the non-linear interaction of multi-phase reactive fluids and elastoplastic solids, inducing high strain-rates and high deformations. 
Each state of matter is governed by a single system of non-linear, inhomogeneous partial differential equations, which are solved simultaneously on the same computational grid, and do not require special treatment of immersed boundaries. 
To this end, the governing equations for solid and reactive multiphase fluid mechanics are written in the same mathematical form and are discretised on a regular Cartesian mesh. 
All phase and material boundaries are treated as diffuse interfaces. 
An interface-steepening technique is employed at material boundaries to keep interfaces sharp whilst maintaining the conservation properties of the system. 
These algorithms are implemented in a highly-parallelised hierarchical adaptive mesh refinement platform, and are verified and validated using numerical and experimental benchmarks. 
Results indicate very good agreement with experiment and an improvement of numerical performance compared to certain existing Eulerian methods, without loss of conservation.

\textcopyright ~British Crown Owned Copyright 2020/AWE
\end{abstract}


\end{frontmatter}

\pagestyle{pprintTitle}
\thispagestyle{pprintTitle}


\section{Introduction}

The ability to model systems containing both solid and reactive-fluid materials is of interest in many industries and academic disciplines.
Examples include applications in explosive safety, mining, explosive welding, and blast-structure interaction.
It is desirable to model materials with different physical properties and equations of motion in a single framework, as this ensures full physical interaction between materials and reduces the complexity of simulations. Furthermore, these problems often feature disparate length scales for which it is desirable to use adaptive mesh refinement (AMR) to better focus computational resources.
Eulerian methods are well suited to problems of this kind, where complex, high strain-rate deformations can result in large topological changes that conventional Lagrangian methods cannot resolve.
However, previous Eulerian methods are not without drawbacks and have challenges associated with their implementation in higher dimensions and with AMR.
Resolving material boundaries in Eulerian methods is a non-trivial problem. This is further exacerbated by materials such as condensed phase explosives that introduce additional phase boundaries between their reactants and products.
This paper outlines a new method that resolves multi-physics and multi-materials simultaneously, and is straightforward to solve on a structured AMR framework.

Most existing Eulerian models capable of simulating fluids and elastoplastic solids are based on tracking material interfaces and fall into one of three broad categories:
\begin{itemize}
\item Methods based on homogenised mixed cells and volume-of-fluid reconstruction, along the lines of the approach described in \citet{benson:1992}. These methods underpin many well-established and legacy multi-material codes.
\item Ghost-fluid methods such as \cite{ortega2015,SchochCoupled,MiNi18,4StatesOfMatter}. Introduced more recently, these methods capture internal boundary conditions by extending the multi-fluid methods from \cite{FedkiwGFM,RiemannGFM,rGFM}. The main drawback of these methods is that they are non-conservative.
\item Cut-cell methods such as \citet{miller:2002} and \citet{barton:2011}. These methods resolve the geometry of cells intersected by interfaces and apply a strict finite volume discretisation.
\end{itemize}
All of these approaches have the advantage that they maintain arbitrarily sharp interfaces, but to do so they involve complex interface reconstructions and mixed-cell algorithms. 
Furthermore, applying load-balancing to these schemes when using AMR is difficult due to the non-uniform numerical methods. Robust implementations can therefore be challenging to construct.
Some of these difficulties have led to the development of hybrid approaches such as Arbitrary Lagrangian Eulerian (ALE) methods and co-simulation methods, based on embedding finite element grids in fluid domains, such as \citet{deiterding2006}. These methods are popular for fluid-structure interaction (FSI) problems, but again introduce the natural complexities of mesh management associated with Lagrangian methods. 

Diffuse interface methods are a practical alternative class of Eulerian techniques. These methods allow a finite-volume computational cell to contain a mixture of several different materials, and are governed by single set of evolution equations that encompasses the physics of all the components in the mixture.
These methods are well established for multi-fluid problems (see for example \citet{Allaire} and \citet{SaurelAbgrall}, and the references therein), but have emerged only recently for coupled solid-fluid dynamics \cite{Barton2019,FavrieElasticDiffuse, FavriePlasticDiffuse, HankExperimental}.
A particular benefit of diffuse interface models is that they provide conservation equations for mass, momentum and energy across interfaces.
Previous work has shown that these schemes have the considerable advantage over interface tracking methods in that they can support genuine fluid mixtures, allowing for the study of phenomena such as cavitation and chemical reaction which rely on physical mixtures \cite{PetitpasCaviation}. This is a major advantage of the scheme at hand, and provides much of the motivation for developing the techniques outlined here. 
In diffuse interface methods the complexities of interface interactions and multi-physics are built into the equations themselves, and only having to solve a single system of evolution equations removes much of the difficulty of constructing numerical methods. 
In the recent model from \citet{Barton2019} for instance, which this work builds upon, the numerical methods are only marginally more complicated than a conventional shock capturing method for inviscid gas dynamics. In short, this means that a computational cell could contain an interface between an arbitrary number of elastoplastic solids, inert fluids, and a reactive mixture, and the numerical method would remain unchanged compared to solving for a single fluid. This approach allows the method to be straightforwardly extended to higher dimensions and adaptive mesh refinement.

This paper extends the diffuse elastoplastic solid method of \citet{Barton2019} to include multi-phase reactive fluids by following along the lines of \citet{MiNi16}. This method will then provide a way to perform simulations that feature elastoplastic solids and reactive fluids in fully-coupled manner, without the need for level sets. 
A practical AMR-based numerical method is detailed, which includes a recently developed interface sharpening technique to counter the diffusion of material interfaces and bring the interface dimensions on par with those of sharp interface methods.
To demonstrate the potential of the method, results are provided for several challenging tests including a multi-dimensional simulation of an explosively formed elastoplastic jet.

\section{Governing Theory}
\label{sec:equations}

The system of equations is based on the \citet{Allaire} five-equation model, augmented with evolution equations for elastoplastic solids following \citet{Barton2019}, and reactive, physically mixing fluids following \citet{MiNi16} (MiNi16). With the base model of conservation equations in place, the additional multi-physics components require closure models to convey specific material behaviour (such as plastic strain rate or reaction rate), which this work will also outline.

\subsection{Evolution equations}
In the interests of focusing on multi-physics, only a summary of the core evolution equations is provided here. Full details, including the derivation, are outlined by \citet{Barton2019}.

Materials are allowed to mix at their interfaces. A material's contribution to a spatially averaged physical quantity is weighted by its volume fraction, $\phi$, in that region. This mixing is referred to as numerical mixing, to distinguish it from the physical mixtures produced by detonations. The state of any material $l$ is characterised by the phasic density $\rho_{(l)}$, volume fraction $\phi_{(l)}$, symmetric left uni-modal stretch tensor $\Vbar$, velocity vector $\mathbf{u}$, and specific internal energy $\mathscr{E}$.
The model assumes mechanical equilibrium; materials in a mixture region share a single velocity, pressure and deviatoric strain. Mixture rules are provided for a consistent definition of thermodynamically averaged quantities in these mixture regions.

The assumption of mechanical equilibrium greatly reduces the number of equations required by the system when multiple materials are considered. The assumption means that only a single velocity and deformation tensor are required, rather than one for each material. This is beneficial in terms of efficiency, by does impose an artificial `stick' boundary condition between solid phases. This issue could be overcome be moving to a non-equilibrium diffuse interface model, but the techniques outlined here for the inclusion of reactive fluids apply to both equilibrium and non-equilibrium models.

For $l=1,\ldots, N$ materials:
\begin{eqnarray}
\frac{\partial \phi_{(l)}}{\partial t} + \frac{\partial \phi_{(l)} u_k}{\partial x_k} &=& \phi_{(l)}\frac{\partial u_k}{\partial x_k}\\
\frac{\partial \rho_{(l)}\phi_{(l)}}{\partial t} + \frac{\partial \rho_{(l)} \phi_{(l)} u_k}{\partial x_k} &=& 0 \\
\frac{\partial \rho u_i }{\partial t} + \frac{\partial (\rho u_i u_k -\sigma_{ik}) }{\partial x_k} &=& 0\\
\frac{\partial \rho E }{\partial t} + \frac{\partial (\rho E u_k - u_i \sigma_{ik}) }{\partial x_k} &=& 0\\
\frac{\partial \Vbar_{ij} }{\partial t} + \frac{\partial \left( \Vbar_{ij}  u_k - \Vbar_{kj} u_i \right) }{\partial x_k} &=& \frac{2}{3}\Vbar_{ij}\frac{\partial u_k}{\partial x_k} - u_i\beta_j - \Phi_{ij} 
\end{eqnarray}
Here $E=\mathscr{E}+|\mathbf{u}|/2$ denotes the specific total energy, $\boldsymbol\sigma$ denotes the Cauchy stress tensor,  $\beta_j=\partial\Vbar_{kj}/\partial x_k$, and $\Phi$ represents the contribution from plastic effects.

Some multi-physics closure models introduce a dependence on material history variables such as the equivalent plastic strain, $\varepsilon_{p(l)}$, or the reaction progress variable, $\lambda_{(l)}$. 
For these variables, additional evolution equations are required:
\begin{eqnarray}
\frac{\partial \rho_{(l)}\phi_{(l)} \alpha_{(l)}}{\partial t} + \frac{\partial \rho_{(l)} \phi_{(l)} \alpha_{(l)} u_k}{\partial x_k} &=& \rho_{(l)}\phi_{(l)}\dot{\alpha}_{(l)} \ .
\end{eqnarray}
Here $\alpha_{(l)}$ represents any such history parameter which is advected and evolved with a material as time progresses. A material may have more than one history variable, in which case $\alpha$ represents a vector.

The system presented here allows for the fully-coupled multi-physics solution of problems involving the interaction of elastoplastic solids with reactive fluid mixtures. The extension to yet more multi-physics applications would follow straightforwardly by the inclusion of additional history parameters and closure relations.

\subsection{Thermodynamics}
\label{sec:thermodynamics}
The internal energy $\mathscr{E}$ for each material is defined by an equation-of-state that conforms to the general form:
\begin{eqnarray}\label{eq_eos_gen}
\mathscr{E}_{(l)}\left(\rho_{(l)},T_{(l)},\dev{\mathbf{H}^e}\right) &= &\mathscr{E}_{(l)}^c\left(\rho_{(l)}\right)+ \mathscr{E}_{(l)}^t\left(\rho_{(l)},T_{(l)}\right) \nonumber \\&& + \mathscr{E}_{(l)}^s\left(\rho_{(l)},\dev{\mathbf{H}^e}\right) \ ,
\end{eqnarray}
where
\begin{equation}
\dev{\mathbf{H}^e} = \ln\left(\Vbar\right) \ 
\end{equation}
is the deviatoric\footnotemark Hencky strain tensor and $T$ is the temperature. The three terms on the right hand side are the contribution due to cold compression or dilation, $\mathscr{E}_{(l)}^c\left(\rho_{(l)}\right)$, the contribution due to temperature deviations, $\mathscr{E}_{(l)}^t\left(\rho_{(l)},T_{(l)}\right)$, and the contribution due to shear strain $\mathscr{E}_{(l)}^s\left(\rho_{(l)},\dev{\mathbf{H}^e}\right)$. The cold compression energy will generally be provided by the specific  closure model for each material, covered in Section \ref{sec:ClosureModels}. The thermal energy is given by
\begin{eqnarray}
\mathscr{E}_{(l)}^t(\rho_{(l)},T)&=& C_{(l)}^{\text{V}}\left(T-T_{(l)}^0\theta_{(l)}^D\left(\rho_{(l)}\right)\right) \ ,
\end{eqnarray}
where $C_{(l)}^{\text{V}}$ is the specific heat capacity, $T_{(l)}^0$ is a reference temperature, and $\theta_{(l)}^{\text{D}}(\rho_{(l)})$ is the non-dimensional Debye temperature.
The Debye temperature is related to the Gr\"uneisen function, $\Gamma(\rho_{(l)})$, via
\begin{equation}
\Gamma_{(l)}(\rho_{(l)}) = \frac{\partial \ln\theta_{(l)}^{\text{D}}(\rho_{(l)})}{\partial \ln(1/\rho_{(l)})} = \frac{\rho_{(l)}}{\theta_{(l)}^{\text{D}}(\rho_{(l)})}\frac{\partial \theta_{(l)}^{\text{D}}(\rho_{(l)})}{\partial\rho_{(l)}} \ .
\end{equation}

The shear energy is given by
\begin{equation}
\mathscr{E}_{(l)}^s(\rho_{(l)},\dev{\mathbf{H}^e}) =  \frac{G_{(l)}\left(\rho_{(l)}\right)}{\rho_{(l)}}  \mathcal{J}^2\left(\dev{\mathbf{H}^e}\right) \ ,
\end{equation}
where $G(\rho)$ is the shear modulus, and
\begin{equation}
\mathcal{J}^2(\dev{\mathbf{H}^e}) = \tr\left(\dev{\mathbf{H}^e}\cdot\dev{\mathbf{H}^e}^{\text{T}}\right)
\end{equation}
is the second invariant of shear strain. This form is chosen such that the resultant stresses are analogous to Hooke's law. \citet{BruhnsStrainEnergy} find that this form provides good empirical agreement for a range of materials and deformations.
\footnotetext{
For any $N\times N$ matrix $\mathbf{M}$, $\dev{\mathbf{M}}:=\mathbf{M}-\frac{1}{N}\tr(\mathbf{M})\mathbf{I}$ denotes the matrix deviator, $\tr(\mathbf{M})$ denotes the trace, and $\mathbf{I}$ denotes the identity matrix.
}

For each component, the Cauchy stress, $\boldsymbol\sigma_{(l)}$, and pressure, $p_{(l)}$, are inferred from the second law of thermodynamics and classical arguments for irreversible elastic deformations:
\begin{align}
\boldsymbol\sigma_{(l)} &= p_{(l)}\mathbf{I} + \dev{\boldsymbol\sigma_{(l)}} \\
p_{(l)} &= \rho^{2}_{(l)} \frac{\partial\mathscr{E}_{(l)}}{\partial \rho_{(l)}} \label{eq:p_energy_derivative}\\ 
\dev{\boldsymbol\sigma_{(l)}} &= 2G_{(l)}\cdot\dev{\mathbf{H}^e} \ .
\end{align}
Although it might appear that the model describes solid materials, inviscid ideal fluids can be considered a special case where the shear modulus is zero, resulting in a spherical stress tensor and no shear energy contribution. It is the equation-of-state for each material that ultimately distinguishes solids from fluids. This formulation lends itself well to diffuse interface modelling where different phases that share the same underlying model can combine consistently in mixture regions.

Mixture rules must be provided to represent the state of regions containing multiple materials in a thermodynamically consistent way. The following mixture rules are applied, following the examples of \citet{MiNi16}, \citet{Allaire} and \citet{Barton2019}:
\begin{align}
1 &= \sum_{l=1}^N\phi_{(l)}\label{vof_mix_rule} \\
\rho &= \sum_{l=1}^N \phi_{(l)} \rho_{(l)} \\
\rho \mathscr{E} &= \sum_{l=1}^N \phi_{(l)} \rho_{(l)} \mathscr{E}_{(l)} \label{Energy_mixture_rule}\\
G &= \frac{\sum_{l=1}^N\left(  \phi_{(l)} G_{(l)}(\rho_{(l)})/\Gamma_{(l)}\right)}{\sum_{l=1}^N\left(\phi_{(l)}/\Gamma_{(l)}\right)}\\
c^2 &= \frac{\sum_{l=1}^N \left( \phi_{(l)} Y_{(l)} c_{(l)}^2 / \Gamma_{(l)} \right)}{\sum_{l=1}^N\left(\phi_{(l)}/\Gamma_{(l)}\right)} \\
\boldsymbol\sigma &= \frac{\sum_{l=1}^N\left(  \phi_{(l)} \boldsymbol\sigma_{(l)}/\Gamma_{(l)}\right)}{\sum_{l=1}^N\left(\phi_{(l)}/\Gamma_{(l)}\right)} \ .
\end{align}
where $c$ is the sound speed and $Y_{(l)} = \frac{\phi_{(l)}\rho_{(l)}}{\rho} $ is the mass fraction.

\subsection{Closure models}
\label{sec:ClosureModels}
It can be seen that, by writing the internal energy in the form outlined, equation \eqref{eq:p_energy_derivative} can be written in the form:
\begin{align}
 p_{(l)} = p_{\text{\scriptsize{ref}},(l)} + \rho_{(l)}\Gamma_{(l)}\left(\mathscr{E}_{(l)} - \mathscr{E}_{\text{\scriptsize{ref}},(l)}\right) \label{eq:MieGruneisenEOS} \ .
\end{align}
Here, $\mathscr{E}_{\text{\scriptsize{ref}},(l)} = \mathscr{E}^c_{(l)} + \mathscr{E}^s_{(l)}$ and $p_{\text{\scriptsize{ref}},(l)} = \rho^{2}_{(l)} \frac{\partial\mathscr{E}_{\text{\scriptsize{ref}},(l)}}{\partial \rho_{(l)}}$. This is the form of the standard Mie-Gr\"uneisen equation-of-state. This encompasses a wide range of different materials, not limited to solids, depending on the choice of the reference curves $\mathscr{E}_{\text{\scriptsize{ref}},(l)}(\rho_{(l)})$, $p_{\text{\scriptsize{ref}},(l)}(\rho_{(l)})$, and $\Gamma_{(l)}(\rho_{(l)})$. These additional freedoms incorporate thermal, compaction and shear effects. For example:
\begin{itemize}
 \item $\mathscr{E}_{\text{\scriptsize{ref}},(l)} = p_{\text{\scriptsize{ref}},(l)} = 0, \ \Gamma(\rho)=\Gamma_0=\gamma-1$ gives the ideal gas law, used for relatively simple gases such as air. Here $\gamma$ is the adiabatic index.
 \item $p_{\text{\scriptsize{ref}},(l)} = -\gamma p_{\infty}, \  \mathscr{E}_{\text{\scriptsize{ref}},(l)} = e_{\infty}, \ \Gamma(\rho)=\Gamma_0=\gamma-1$ gives the stiffened gas equation of state, used to model denser fluids such as water. 
 \item $\Gamma(\rho)=\Gamma_0=\gamma-1$ and 
 \begin{align}
   p_{\text{\scriptsize{ref}},(l)}(\rho_{(l)}) &= {\cal A}e^{-{\cal R}_1\frac{\rho_0}{\rho_{(l)}}}+{\cal B}e^{-{\cal R}_2\frac{\rho_0}{\rho_{(l)}}} \\
   \mathscr{E}_{\text{\scriptsize{ref}},(l)}(\rho_{(l)}) &= \frac{{\cal A}}{{\cal R}_1\rho_0}e^{-{\cal R}_1\frac{\rho_0}{\rho_{(l)}}}+\frac{{\cal B}}{{\cal R}_2\rho_0}e^{-{\cal R}_2\frac{\rho_0}{\rho_{(l)}}}
 \end{align}
 gives the JWL (Jones--Wilkins--Lee) equation-of-state, widely used for condensed phase explosives or reaction products \cite{MiNi16}.
 \item $\Gamma(\rho)=\Gamma_0=\gamma-1$ and
 \begin{align}
   \mathscr{E}_{\text{\scriptsize{ref}},(l)}(\rho_{(l)}) &= \frac{K_0}{2\rho_{(l)}\bar{\alpha}^2}\left(\left(\frac{\rho_{(l)}}{\rho_0}\right)^{\bar{\alpha}}-1\right) + \mathscr{E}^s_{(l)} \\
   G(\rho_{(l)}) &= G_0\left(\frac{\rho_{(l)}}{\rho_0}\right)^{\bar{\beta}+1} \\
   p_{\text{\scriptsize{ref}},(l)}(\rho_{(l)}) &= \rho^{2}_{(l)} \frac{\partial\mathscr{E}_{\text{\scriptsize{ref}},(l)}}{\partial \rho_{(l)}}
 \end{align}
 gives the Romenskii equation-of-state, used for elastic solids \cite{RomenskiiEOS}. Here, $\bar{\alpha}$ and $\bar{\beta}$ are material dependent parameters, $K_0$ and $G_0$ are the reference bulk and shear moduli and $\rho_0$ is the reference density.
\end{itemize}

\subsubsection{Reactive Fluids}
The reactive components considered in this work do not model the specific chemistry of any given reaction, rather the approach will be to model the effects on the continuum scale. It is assumed for simplicity that reactive fluids are physical mixtures\footnote{This reactive mixture is referred to as a physical mixture to distinguish it from the numerical mixtures inherent in diffuse interface schemes. This distinction is drawn as both the origin (physical reactive source terms) and the length scale of the mixing are different.} composed of two components: the reactant, $\alpha$, and the product $\beta$. Each of these components may be governed by a different equation-of-state with different parameters. The reaction will be tracked by following the reaction progress variable, $\lambda$; when only reactants are present $\lambda=1$, and when $\lambda=0$ the reaction has fully converted reactants to products. It will be assumed that reactants will turn to products in a simple one-step exothermic reaction. The following equation is then added to the system of equations:

\begin{align}
\frac{\partial \rho_{(l)}\phi_{(l)} \lambda_{(l)}}{\partial t} + \frac{\partial \rho_{(l)} \phi_{(l)} \lambda_{(l)} u_k}{\partial x_k} &= \rho_{(l)}\phi_{(l)}\dot{\lambda}_{(l)} \ .
\end{align}

The rate at which the reaction occurs, $\dot{\lambda}$, is a closure model defined by the choice of reaction rate law. This term is then included as a source term for the reaction progress variable. Specific examples of the reaction rates are given in Section \ref{sec:Validation}.

Following \citet{MiNi16}, reactive fluids are modelled as a mixture, with this mixture being treated as a single material for the purposes of bookkeeping. In other words, the system of equations evolves the partial density of the combined mixture ($\alpha_{(l)}\rho_{(l)}$), the volume fraction occupied by both components of the mixture ($\alpha_{(l)})$ and the reaction progress variable ($\lambda_{(l)}$), rather than evolving the partial densities of each phase in the mixture. However, the partial densities of each component in the mixture will still be required when evaluating the equation-of-state. To this end, a root finding procedure is detailed in \ref{app:RootFinding}.

The energy input generated by the reaction will arise by including a reaction energy as a term in the energy of the products:
\begin{align}
\mathscr{E}_{\text{\scriptsize{ref}},(l)} \rightarrow \mathscr{E}_{\text{\scriptsize{ref}},(l)} - Q \ , 
\end{align}
where $Q$ is the energy released by the reaction. It is this energy input which will drive self-sustaining detonation waves.
Mixture rules for mass fraction and internal energy must be provided to relate how the two components combine in mixture regions \cite{MiNi16}:
\begin{eqnarray}
\frac{1}{\rho_{(l)}} &=& \frac{\lambda_{(l)}}{\rho_{\alpha}} + \frac{1-\lambda_{(l)}}{\rho_{\beta}}\label{eq:rho_physicalmixturerule}\\
\mathscr{E}_{(l)} &=& \lambda_{(l)} \mathscr{E}_{\alpha}+(1-\lambda_{(l)})\mathscr{E}_{\beta} \label{eq:E_physicalmixturerule} \ .
\end{eqnarray}

The sound speed in a physical mixture is calculated using the reaction-progress-variable-weighted form from \citet{MiNi16}:
\begin{subequations}
\begin{align}
 c_{\mbox{\scriptsize{mix}},(l)}^2 &= \frac{\frac{p}{\rho_{(l)}^2} -\left(\pdv{\mathscr{E}_{(l)}}{\rho_{(l)}}\right)_{p}}{\left(\pdv{\mathscr{E}_{(l)}}{p}\right)_{\rho_{(l)}}} \label{eq:mixtureSoundSpeed} \\
\left(\pdv{\mathscr{E}_{(l)}}{p}\right)_{\rho_{(l)}} &= \lambda_{(l)} \left(\pdv{\mathscr{E}_{\alpha}}{p}\right)_{\rho_{(l)}}+ (1-\lambda)\left(\pdv{\mathscr{E}_{\beta}}{p}\right)_{\rho_{(l)}} \\
\left(\pdv{\mathscr{E}_{(l)}}{\rho_{(l)}}\right)_{p} &= \lambda_{(l)} \left(\pdv{\mathscr{E}_{\alpha}}{\rho_{\alpha}}\right)_{p}\left(\pdv{\rho_{\alpha}}{\rho_{(l)}}\right)_{p}+  (1-\lambda) \left(\pdv{\mathscr{E}_{\beta}}{\rho_{\beta}}\right)_{p}\left(\pdv{\rho_{\beta}}{\rho_{(l)}}\right)_{p} \ .
\end{align}
\end{subequations}

\subsubsection{Plasticity}

The introduction of plasticity through the source term $\Phi$ follows the method of convex potentials (see for example \citet{Ottosen2005}) and is therefore thermodynamically compatible.
In this approach, the von Mises yield criterion forms the scalar potential which leads to the plastic flow rule:
\begin{align}
 \Phi &= \chi \sqrt{\frac{3}{2}}\frac{\dev{\sigma}}{\Jnorm{\dev{\sigma}}} \Vbar \ .
\end{align}
The plastic flow rate $\chi$ is a closure model and must be suitable for arbitrary mixtures. For a mixture of $N$ materials:
\begin{align}
\chi = \frac{\sum_{l=1}^N\left(  \phi_{(l)} \chi_{(l)}(\rho_{(l)})/\Gamma_{(l)}\right)}{\sum_{l=1}^N\left(\phi_{(l)}/\Gamma_{(l)}\right)} \ .
\end{align}
If any material is does not obey a plasticity model, that material contributes $\chi_{(l)} = 0$.

The form of $\chi$ relates the particular material flow model. This paper considers both ideal plasticity, where $\chi$ is a Heaviside function such that the relaxation is non-zero only when the stress exceeds the yield surface $\sigma_Y$:
\begin{align}
 \chi_{(l)} = \chi_{(l)}^0 H\left[ \sqrt{\frac{3}{2}} ||\text{dev}({\boldsymbol\sigma})_{(l)}|| -\sigma_Y \right] \ ,
\end{align}
and the rate sensitive isotropic work-hardening plasticity outlined by \citet{JohnsonCook}:
\begin{equation}
\chi_{(l)} = \chi_{(l)}^0 \exp\left[\frac{1}{c_3}\left(\frac{\sqrt{\frac{3}{2}}||\dev{{\boldsymbol\sigma}_{(l)}}||}{\sigma_Y\left(\varepsilon_{p,(l)}\right)}-1\right)\right], \label{JCPlasticity}
\end{equation}
where ${\chi}_{0}>0$ is the reference plastic strain-rate and the constant $c_3$ controls the rate dependency. In the Johnson and Cook model the yield stress is given by:
\begin{align}
 \sigma_Y\left(\varepsilon_{p,(l)}\right) = \left(c_1+c_2(\varepsilon_{p,(l)})^n\right)\left( 1 - \left(\frac{T-T_0}{T_{\text{melt}}-T_0}\right)^m   \right)\ ,
\end{align}
where $c_1$ is the yield stress, $c_2$ is the strain hardening factor, $n$ is the strain hardening exponent, $T_{\text{melt}}$ is the melting temperature of the material, $T_0 = 298$ K is a reference temperature, and $m$ is the thermal softening exponent. This yield surface is a function of the accumulated plastic strain, $\varepsilon_p$, making it necessary to add an additional evolution equation to the system to advect and evolve the plastic strain:
\begin{align}
\frac{\partial \rho_{(l)}\phi_{(l)} \varepsilon_{(l),p}}{\partial t} + \frac{\partial \rho_{(l)} \phi_{(l)} \varepsilon_{(l),p} u_k}{\partial x_k} = \rho_{(l)}\phi_{(l)}\chi_{(l)} \ .
\end{align}
Using the Johnson-Cook model in this way results in a viscoplastic flow rule, where plastic deformations can accumulate from the onset of loading. Note however that the parameter $c_3$ is usually small such that $\chi_{(l)}\ll\chi_{(l)}^0$ for stresses much below the characteristic stress.
Indeed, as $c_3\rightarrow0$ the plastic flow becomes rate independent and the stress becomes bounded by a yield surface.

\section{Numerical Approach}

The model is solved on a Cartesian mesh with local resolution adaptation in space and time. This is achieved using the AMReX software from Lawrence Berkely National Laboratory \cite{amrex}, which includes an implementation of the structured adaptive mesh refinement (SAMR) method of \citet{berger:1988} for solving hyperbolic systems of partial differential equations (PDEs) of the form of equation \eqref{eq:sys_vec_form}.
In this approach, cells of identical resolution are grouped into logically rectangular sub-grids or `patches'. 
Refined grids are derived recursively from coarser ones, based upon a flagging criterion, to form a hierarchy of successively embedded levels.
All mesh widths on level $l$ are $r_l$-times finer than on level $l-1$, i.e. $\Delta t_l:=\Delta t_{l-1}/r_l$ and $\Delta \mathbf{x}_{l}:=\Delta \mathbf{x}_{l-1}/r_l$ with $r_l\in\mathbb{N}, r_l\ge 2$ for $l>0$ and $r_0=1$. 
The numerical scheme is applied on level $l$ by calling a single-grid update routine in a loop over all patches constituting the level. 
The discretisation of the constitutive models does not differ between patches or levels, so for clarity the method shall be described for a single sub-grid.
Cell centres are denoted by the indices $i,j,k\in\mathbb{Z}$ and each cell $C^{l}_{ijk}$ has the dimensions $\Delta \mathbf{x}^l_{ijk}$.

The system of equations can be written compactly in vector form by separating it into various qualitatively different parts: a conservative hyperbolic part for each spatial dimension, non-conservative terms from the volume fraction and stretch tensor updates, a source term due to plastic flow, a source term due to reactive species, and a source term to account for geometrical effects. This can be written as:
\begin{equation}
 \frac{\partial \mathbf{q}}{\partial t}+ \frac{\partial \mathbf{g}_k}{\partial x_k} =  \mathbf{s}_{\text{\scriptsize{non-con.}}} + \mathbf{s}_p + \mathbf{s}_r + \mathbf{s}_g \ . \label{eq:sys_vec_form} 
\end{equation}
Subject to the closure relations previously outlined, this is given by:
\begin{eqnarray}
 \pdv{t}\mqty(	\phi_{(l)} \\
		\phi_{(l)}\rho_{(l)} \\
		\phi_{(l)}\rho_{(l)}\lambda_{(l)} \\
		\phi_{(l)}\rho_{(l)}\varepsilon_{p,(l)} \\
		\rho u_i \\
		\rho E \\
		\Vbar_{ij} \\)  
		+ \pdv{x_k}\mqty(\phi_{(l)}u_k \\
		\phi_{(l)}\rho_{(l)}u_k \\
		\phi_{(l)}\rho_{(l)}\lambda_{(l)}u_k \\
		\phi_{(l)}\rho_{(l)}\varepsilon_{p,(l)}u_k \\
		\rho u_iu_k -\sigma_{ik} \\
		\rho Eu_k - u_i\sigma_{ik} \\
		\Vbar_{ij}u_k - \Vbar_{kj}u_i \\) + \cdots \\
		= \mqty(\phi_{(l)}\pdv{u_k}{x_k} \\
		0 \\
		0 \\
		0 \\
		0 \\
		0 \\
		\frac{2}{3}\Vbar_{ij}\pdv{u_k}{x_k} - u_i\pdv{\Vbar_{kj}}{x_k} \\)
		+ \mqty(0 \\
			0 \\
			0 \\
			\phi_{(l)}\rho_{(l)}\dot{\varepsilon}_{p,(l)} \\
			0 \\
			0 \\
			\Phi_{ij} \\)
		+ \mqty(0 \\
			0 \\
			\phi_{(l)}\rho_{(l)}\dot{\lambda}_{(l)} \\
			0 \\
			0 \\
			0 \\
			0 \\)  + \mathbf{s}_g
\end{eqnarray}
When considering cylindrical symmetry, the conservative variables and geometrical term are given by:
\begin{align}
 \vb{q} =\mqty(	\phi_{(l)} \\
		\phi_{(l)}\rho_{(l)} \\
		\phi_{(l)}\rho_{(l)}\lambda_{(l)} \\
		\phi_{(l)}\rho_{(l)}\varepsilon_{p,(l)} \\
		\rho u_r \\
		\rho u_z \\
		\rho E \\
		\Vbar_{ij} \\),		
\ \  \vb{s}_g =-\frac{1}{r} \mqty( 0 \\
		\phi_{(l)}\rho_{(l)} u_r \\
		\phi_{(l)}\rho_{(l)}\lambda_{(l)} u_r \\
		\phi_{(l)}\rho_{(l)}\varepsilon_{p,(l)} u_r \\
		\rho u_r^2 - \sigma_{rr} + \sigma_{\theta\theta} \\
		\rho u_zu_r -\sigma_{rz}\\
		\rho Eu_r-(u_r\sigma_{rr} + u_z\sigma_{zr}) \\
		\frac{1}{3} \Vbar_{ij} u_r - \delta_{i\theta}\Vbar_{ij} u_r ) \ .
\end{align}

The inhomogeneous system is integrated for time intervals $[t^n,t^{n+1}]$ where the time-step $\Delta t=t^{n+1}-t^n$ is chosen to be a fraction of the global maximum allowable time step required for stability of the hyperbolic update method.
For high strain-rate or highly reactive applications, the plastic relaxation or reactions can occur over smaller time scales, which can lead to local stiffness.
To address this issue, without resorting to forecasting stiff zones and resolving the time scales of irreversible physical processes, Godunov's method of fractional steps is used, where the hyperbolic part is updated first, followed by serially adding in contributions from each source term:
\begin{eqnarray}
\frac{\partial \mathbf{q}}{\partial t}  &=& -\frac{\partial \mathbf{g_k}}{\partial x_k} + \mathbf{s}_{\text{\scriptsize{non-con.}}} \qquad  \text{IC:} \;\;\mathbf{q}^n  \xrightarrow \Delta t {\mathbf{q}}^{\mbox{\ding{73}}} \label{exp_update}\\
\frac{\partial \mathbf{q}}{\partial t}  &=& \mathbf{s}_x \qquad\qquad\qquad\;\;\; \text{IC:} \;\;\mathbf{q}^{\mbox{\ding{73}}} \xrightarrow{\Delta t} {\mathbf{q}}^\bigstar\label{imp_udate_p} \ ,
\end{eqnarray}
where the result of the each step is used as the initial condition (IC) for the next.
Here $\vb{s}_x=\vb{s}_p, \vb{s}_r, \vb{s}_g$. Once all source terms have been added, the last $\vb{q}^{\bigstar}$ becomes $\vb{q}^{n+1}$.

\subsection{Hyperbolic update}

Employing the method of lines and replacing the spatial derivatives with a conservative approximation, the hyperbolic system can be written
\begin{equation}\label{sys_mat_dis}
 \frac{\text{d}}{\text{d}t}{\mathbf{q}}_{ijk}+\mathcal{D}_{ijk}\left({\mathbf{q}}\right) = 0,
\end{equation}
where ${\mathbf{q}}_{ijk}$ represents the vector of conservative variables stored at cell centres,  and 
\begin{eqnarray}
 \mathcal{D}_{ijk} :=&& \frac{1}{\Delta x^1_{ijk}}\left(\widetilde{\mathbf{g}}^1_{i+1/2,jk}-\widetilde{\mathbf{g}}^1_{i-1/2,jk}\right)\nonumber\\
&+&\frac{1}{\Delta x^2_{ijk}}\left(\widetilde{\mathbf{g}}^2_{i,j+1/2,k}-\widetilde{\mathbf{g}}^2_{i,j-1/2,k}\right)\nonumber\\
&+&\frac{1}{\Delta x^3_{ijk}}\left(\widetilde{\mathbf{g}}^3_{ij,k+1/2}-\widetilde{\mathbf{g}}^3_{ij,k-1/2}\right) - \mathbf{s}_{\text{\scriptsize{non-con.}},ijk},\label{eq:spat_op}
\end{eqnarray}
where $\widetilde{\mathbf{g}}^l_{m\pm1/2}$, for $m=i,j,k$, are the cell wall numerical flux functions. 
The numerical fluxes are computed through successive sweeps of each spatial dimension and summed according to equation \eqref{eq:spat_op}. Fluxes are computed using the HLLD solver from \citet{Barton2019}. 
To achieve higher order spatial accuracy, the initial conditions for the Riemann solver are taken to be MUSCL reconstruction of cell centred primitive variables. It should be noted that this procedure is generally carried out on the conservative variables, but this work instead follows \citet{JohnsenColonius} who showed that, for multi-material problems, reconstruction of the primitive variables leads to fewer oscillations around interfaces. An artificial interface reconstruction is applied to reduce numerical diffusion around interfaces. This is achieved using the Tangent of Hyperbola INterface Capturing (THINC) method: an algebraic interface reconstruction technique that fits a hyperbolic tangent function to variables inside a cell. 
In contrast to \citet{Barton2019} who used the original THINC method of \citet{XiaoTHINC}, the more recent MUSCL-BVD-THINC scheme of \citet{BVDTHINC} is employed. This more recent scheme provides an additional check to minimise oscillations by comparing the reconstructed state's cell boundary variation with the previously calculated MUSCL reconstruction. THINC-reconstructed states are only accepted when their total boundary variation is lower than that of the MUSCL scheme alone. This algorithm is provided in \ref{app:THINC} for convenience.
The non-conservative source terms that result from the volume fraction and deformation tensor updates are added in the hyperbolic step following a procedure similar to that used by \citet{MiNi16}, \citet{Allaire}, and \citet{Barton2019}.
To achieve a higher temporal resolution in the update of the hyperbolic terms, a third order Runge-Kutta time integration is used:
\begin{align}\label{rk}
\mathbf{q}^{(1)}    &= \mathbf{q}^{n} - \Delta t  \mathcal{D}\left({\mathbf{q}}^{n}\right)\\
\mathbf{q}^{(2)}    &= \mathbf{q}^{(1)} - \Delta t \mathcal{D}\left({\mathbf{q}}^{(1)}\right)\\
\mathbf{q}^{(3)} &= \frac{3}{4} \mathbf{q}^{n} + \frac{1}{4} \mathbf{q}^{(2)} \\
\mathbf{q}^{(4)}    &= \mathbf{q}^{(3)} - \Delta t  \mathcal{D}\left({\mathbf{q}}^{(3)}\right)\\
\mathbf{q}^{\mbox{\ding{73}}} &= \frac{1}{3} \mathbf{q}^{n} + \frac{2}{3} \mathbf{q}^{(4)} \ .
\end{align}

\subsection{Plastic update}
\label{sec:numericalPlastic}
The plastic update is performed after the hyperbolic step as detailed above. The potentially stiff ODEs governing the plasticity evolution are solved using an analytical technique as detailed by \citet{Barton2019}, which reduces the problem to a single ODE for each material in a given cell. The algorithm can be summarised as follows:
\begin{align}
\left(\Vbar\right)^{\bigstar} &= \exp\left(\dev{\mathbf{H}^e}^{\bigstar}\right) \label{eq:eq_plast_qnew} \\
\dev{\mathbf{H}^e}^{\bigstar} &= \frac{\mathcal{J}^{\bigstar}}{\mathcal{J}^{\mbox{\ding{73}}}} \dev{\mathbf{H}^e\left({\Vbar}^{\mbox{\ding{73}}}\right)}\label{eq:eq_plast_hnew} \\
\mathcal{J}^{\bigstar} &= \frac{\sum_{l=1}^N \left(\phi_{(l)}/\Gamma_{(l)}\right)\mathcal{J}^{\bigstar}_{(l)}}{\sum_{l=1}^N \phi_{(l)}/\Gamma_{(l)}} \label{eq:eq_plast_jnew} \\
\mathscr{R}\left(\mathcal{J}_{(l)}^{\bigstar}\right) &= \mathcal{J}_{(l)}^{\bigstar} - \mathcal{J}_{(l)}^{\mbox{\ding{73}}} + \Delta t \chi_{(l)}\left(\mathcal{J}_{(l)}^{\bigstar},\varepsilon_{p_{(l)}}^{\bigstar}\left(\mathcal{J}_{(l)}^{\bigstar}\right)\right) \label{eq:eq_plast_ojb}
\end{align}

The last objective equation is solved using a simple bisection algorithm between the limits $\mathcal{J}_{(l)}^{\bigstar}\in[0:\mathcal{J}_{(l)}^{\mbox{\ding{73}}}]$.

It is necessary to evaluate this algorithm, at least in part, everywhere in a problem irrespective of whether a material adheres to a plasticity law or not, since the algorithm ensures that the stretch tensor remains symmetric, and the condition $\Vbar=\mathbf{I}$ is true for fluids. In essence, the plastic algorithm also applies the condition:
\begin{equation}\label{eq:eq_symm_rein}
\Vbar \leftarrow \sqrt{\Vbar{\Vbar}^{\text{T}}} \ ,
\end{equation}
ensuring the unimodular and symmetry properties of the tensor are retained, which may not have been the case after the hyperbolic step.
The implementation of this algorithm by \citet{Barton2019} employs singular value decomposition (SVD) so that equation \eqref{eq:eq_plast_qnew} and equation \eqref{eq:eq_plast_hnew} are evaluated exactly. 
Since SVD is relatively expensive, this part of the overall numerical scheme constitutes a primary overhead. To alleviate this computational burden, the following modifications are made.

The logarithmic strain tensor in equation \eqref{eq:eq_plast_hnew} is evaluated using a variant of the approximation by \citet{Bazant1998}:
\begin{equation}
\dev{\mathbf{H}^{e}} \approx \frac{1}{2}\mathbf{H}_B^{e^{(1)}}\left(\Vbar{\Vbar}^{\text{T}}\right) \ ,
\end{equation}
where
\begin{equation}
\mathbf{H}_B^{e^{(m)}}\left(\Vbar\right) = \frac{1}{2m}\left({\Vbar}^m-{\Vbar}^{-m}\right) \ ,
\end{equation}
and the invariants evaluated from this in turn; the exponential matrix in equation \eqref{eq:eq_plast_qnew} is evaluated using the first Pad\'{e} approximant:
\begin{equation}
(\Vbar)^{n+1} \approx  \left( 1- \frac{1}{2}\dev{\mathbf{H}^{e}}^{n+1}\right)^{-1}\left(1 + \frac{1}{2}\dev{\mathbf{H}^{e}}^{n+1}\right) \ .
\end{equation}
The approximation of the strain was proposed in \citet{Barton2019} but was used only where stresses are computed for the numerical flux functions. 
Use of the first Pad\'{e} approximate is found to be sufficient over second or higher variants since the second norm of $\mathbf{H}^e$ is known from the outset not to exceed sufficiently small values for the materials of interest.
These approximations avoid the use of a SVD, and the algorithm reduces to simple matrix operations.
Further efficiency can be achieved noting $\Vbar$ and $\dev{\mathbf{H}^e}$ are unimodular, meaning their inverses are equivalent to their cofactor, the evaluation of which avoids computing the determinant.

\subsection{Reactive update}

The potentially stiff reactive update consists of solving the equation:
\begin{eqnarray}
\dv{t} \left(\rho_{(l)}\phi_{(l)} \lambda_{(l)}\right) = \rho_{(l)}\phi_{(l)}\dot{\lambda}_{(l)} \ .
\end{eqnarray}
The fourth order RK4 integration method was used for this task.

In the course of computation it is necessary to evaluate the equations-of-state for reactive mixtures, and to be able to convert between the primitive variables and the conservative variables. For this, it is required to find the partial densities of the reactive mixture components, $\rho_{\alpha}$ and $\rho_{\beta}$. As only the total mixture density is evolved, a non-linear root finding procedure is required. The outline of this root finding procedure is included in \ref{app:RootFinding}. In this work, only a single physical mixture species is considered in a simulation, as including more than this requires a multi-dimensional root finding procedure, which has not yet been implemented. Once these component densities are defined, equation \eqref{eq:E_physicalmixturerule} may be combined with the equation of state for each material \eqref{eq:MieGruneisenEOS} to evaluate the mixture rules.

\begin{table*}
\begin{center}

\begin{tabular}{|c|c|c|c|c|c|c|}
\hline
Material & $\rho_0$ /kg$\cdot$m$^{-3}$ & $K_0$ / GPa & $G_0$ / GPa & $\Gamma$ & $\bar{\alpha}$ & $\bar{\beta}$ \\
\hline
Copper & 8930 & 136.5 & 39.4 & 2 & 1 & 3 \\
\hline
\end{tabular} 
\caption{The Romenskii equation-of-state parameters for copper, taken from \citet{Barton2019}.}~\\[0.2cm]
\label{tab:RomenskiiCopper}

\begin{tabular}{|c|c|c|c|c|c|}
\hline
Material & $\rho_0$ /kg$\cdot$m$^{-3}$ & $\gamma$ & $p_{\infty}$ /GPa & $C^V$ /J$\cdot$kg$^{-1}$K$^{-1}$& $Q$ /MJ$\cdot$kg$^{-1}$ \\
\hline
Reactant & 1600 & 4.0 & 1.0 & 2000 & -\\
Product  & -    & 3.0 & 0.0 & 1000 & 3.68\\
\hline
\end{tabular} 
\caption{The generic multiphase taken from \citet{SchochCoupled} and \citet{PetitpasExplsosive}.}~\\[0.2cm]
\label{tab:genericExplosive}

\begin{tabular}{|c|c|c|c|c|c|c|c|c|}
\hline
Material & $\rho_0$ /kg$\cdot$m$^{-3}$ & ${\cal A}$ / 10$^{11}$ Pa & ${\cal B}$ / 10$^{11}$ Pa & ${\cal R}_1$ & ${\cal R}_2$ & $\Gamma$ & $C^V$ /J$\cdot$kg$^{-1}$K$^{-1}$& $Q$ /MJ$\cdot$kg$^{-1}$ \\
\hline
Reactant & 1905 & 778.1 & -0.0503 & 11.3 & 1.13 & 0.8938 & 1305.5 & 0 \\
Product  & -    & 14.81 & 0.6379 & 6.2 & 2.2 & 0.5 & 524.9 & 3.622 \\
\hline
\end{tabular} 
\caption{The JWL equation of state parameters for LX-17, taken from \cite{TarverLX17CornerTurning, TarverLX17HockeyPuck, IoannouLX17CornerTurning}.}~\\[0.2cm]
\label{tab:JWL}

\begin{tabular}{|c|c|c|c|c|c|c|c|c|c|c|c|}
\hline
$a$ & $b$ & $c$ & $d$ & $e$ & $g$ & $x$ & $y$ & $z$ & $F_{ig}$ & $F_{G_1}$ & $F_{G_2}$ \\
\hline
0.22 & 0.667 & 0.667 & 1 & 0.667 & 0.667 & 7 & 3 & 1 & 0.02 & 0.8 & 0.8 \\
\hline
\end{tabular}~\\[0.1cm]
\begin{tabular}{|c|c|c|}
\hline
I /s$^{-1}$ & G$_1$ / ($10^{11}$ Pa)$^{-y}$ s$^{-1}$   & G$_2$ / ($10^{11}$ Pa)$^{-z}$ s$^{-1}$ \\
\hline
$4\times 10^{12}$& 4500 $\times 10^{6}$ & 30 $\times 10^{6}$ \\
\hline
\end{tabular}
\caption{The ignition and growth reaction rate parameters for LX-17.}~\\[0.2cm]
\label{tab:IandG}

\end{center}
\end{table*}

\section{Validation and Evaluation}
\label{sec:Validation}
\subsection{One Dimensional Test}
\label{sec:1DTest}

The interaction of reactive fluids with elastic solids was examined to ensure this multiphase aspect can be handled by the scheme. A one-dimensional test employed by \citet{SchochCoupled} was considered. This test involves a detonation wave travelling through a multiphase explosive and then hitting a slab of purely elastic copper. The domain spans $x=[0,0.4]$ m, a CFL = 0.9 was used, and a base resolution of 1000 cells with 2 layers of AMR was employed, giving and effective resolution of and $N=4000$ to match \citet{SchochCoupled}. The initial conditions are quiescent copper occupying the region $x<0.15$ m, ambient explosive occupying the range $0.15<x<0.35$, and a booster region of high pressure explosive at  $0.35<x < 0.4$. The solid is governed by the Romenskii equation-of-state, with the parameters given in Table \ref{tab:RomenskiiCopper}. The explosive used is the idealised multiphase mixture used by \citet{SchochCoupled}, taken from \citet{PetitpasExplsosive}, given in Table \ref{tab:genericExplosive}. The reaction rate law used by \citet{SchochCoupled} is a simplified reaction-progress-variable-based law. In terms of this model, this law can be expressed as:
\begin{eqnarray}
 \dot{\lambda} = k \sqrt{\lambda} \ , 
\end{eqnarray}
where $k = 2 \times 10^6$ s$^{-1}$. The reaction is switched on when the pressure exceeds 10$^8$ Pa. To ignite the detonation wave, the booster region $0.35 < x < 0.40$ m has a pressure of 10$^9$ Pa, with the rest of the domain having a pressure of 10$^5$ Pa.

\begin{figure}
 \centering
 \includegraphics[width = 0.5\textwidth]{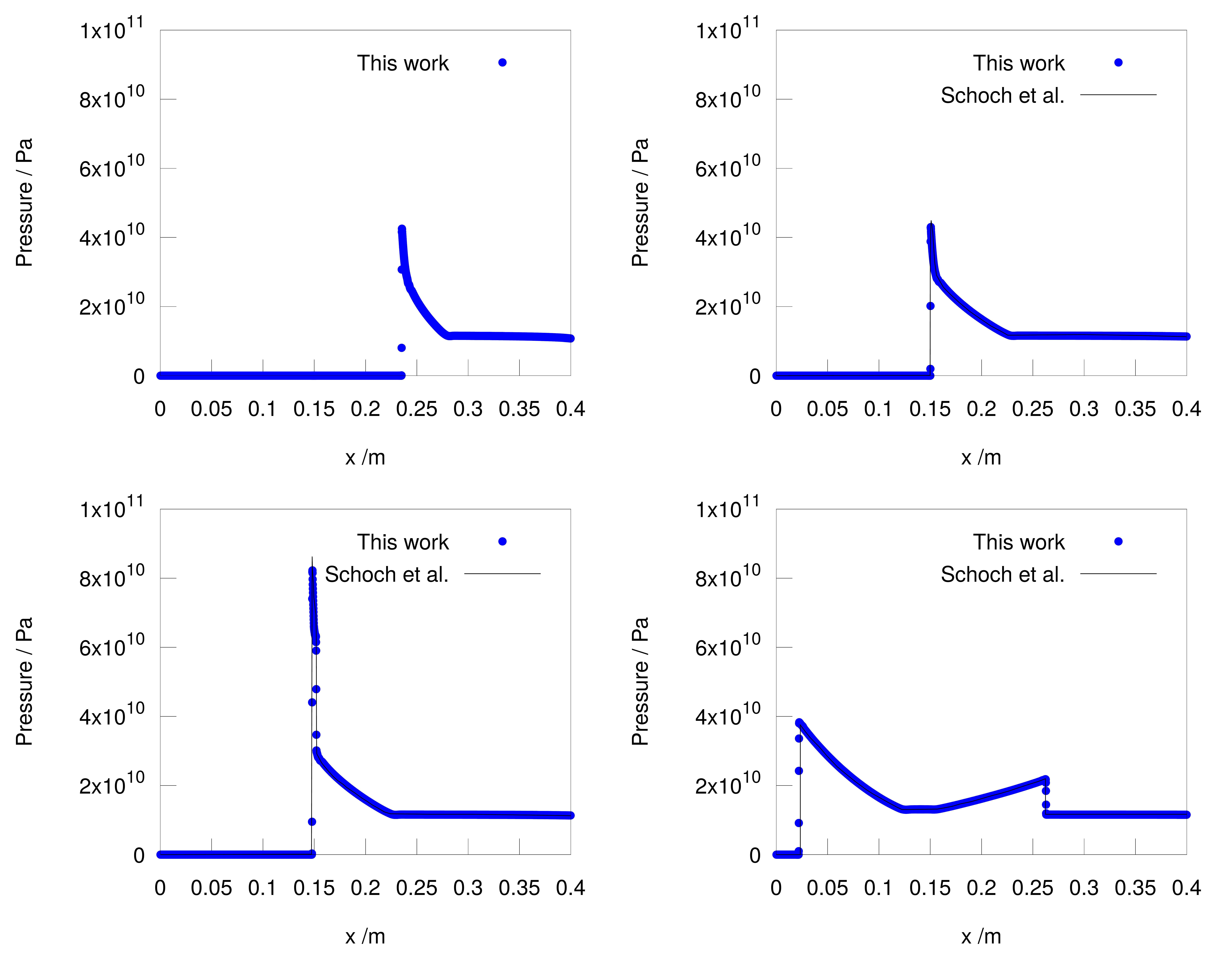}
 \caption{A detonation wave impacting an elastoplastic solid, transmitting a shock wave. This test is compared to the sharp interface results of \citet{SchochCoupled}, showing excellent agreement. The images are taken at t = -10, 0, 0.4, 19.5 $\mu$s after the collision of the detonation wave at the interface, and show the pressure given as $-\sigma_{xx}$. This test demonstrates the ability of the model to simultaneously handle reactive fluid mixtures and elastic solids without the need for level sets.}
 \label{fig:Schoch3}
\end{figure}

Figure \ref{fig:Schoch3} shows this test over time. Initially a detonation wave forms, travelling from right to left, which then collides with the copper interface. After collision, a shock is transmitted into the copper and a reflected wave travels back into the reaction products. A correct, stable interaction between the fluid detonation wave and the quiescent copper was observed, matching the overlaid sharp interface results of \citet{SchochCoupled}.

\begin{figure*}
 \centering
 \includegraphics[width = 0.85\textwidth]{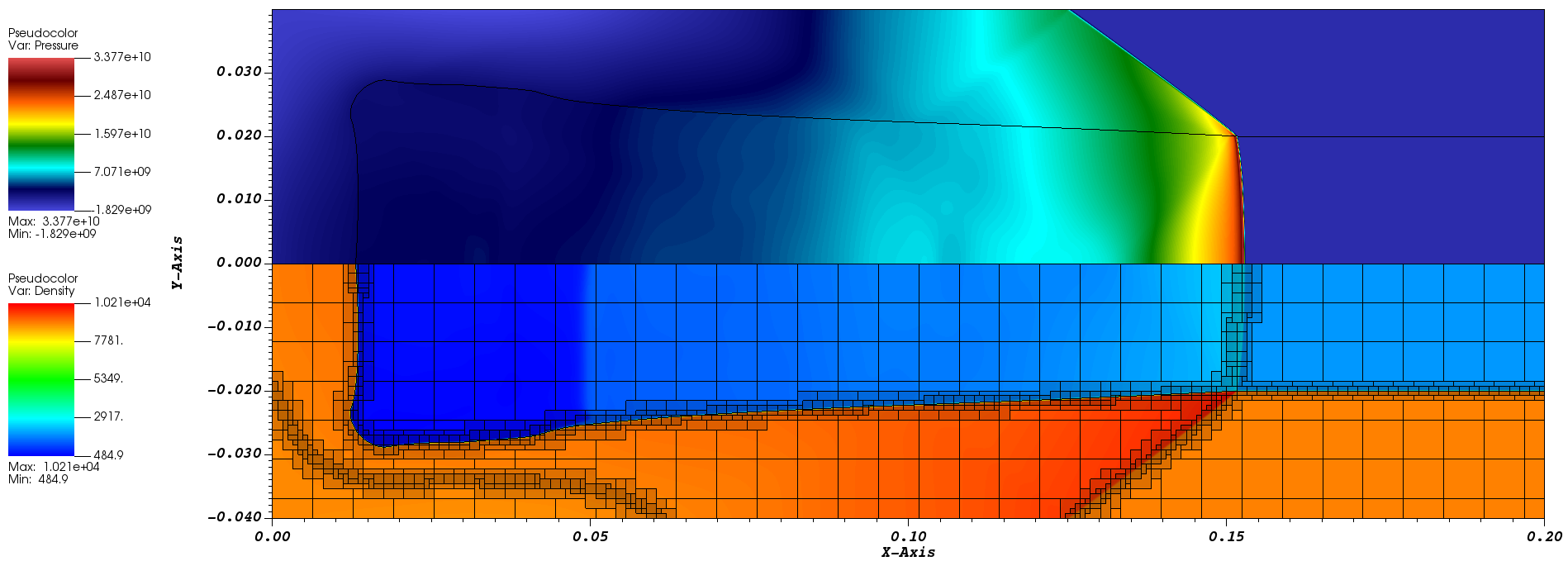}
 \includegraphics[width = 0.85\textwidth]{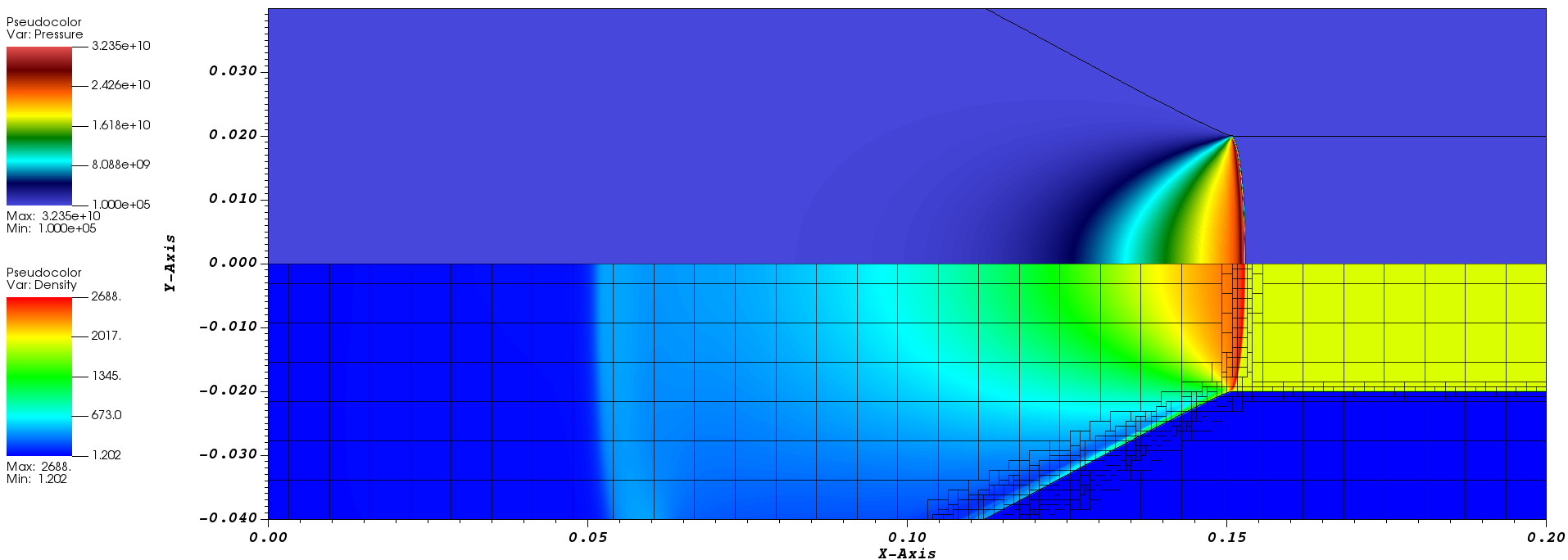}
 \caption{The 20 mm radius LX-17 rate stick test, showing the confined (\textit{Top panel}) and unconfined cases (\textit{Bottom panel}). Each image depicts pressure in Pa (\textit{Upper half}) and density in kg$\cdot$m$^{-3}$ with overlaid AMR levels (\textit{Lower half}). These images are taken at 15 $\mu$s. This test demonstrates the dual multi-physics capabilities of elastoplastic solids and reactive fluids working simultaneously.}
  \label{fig:ratestick}
\end{figure*}

\begin{figure}
 \centering
 \includegraphics[width = 0.5\textwidth]{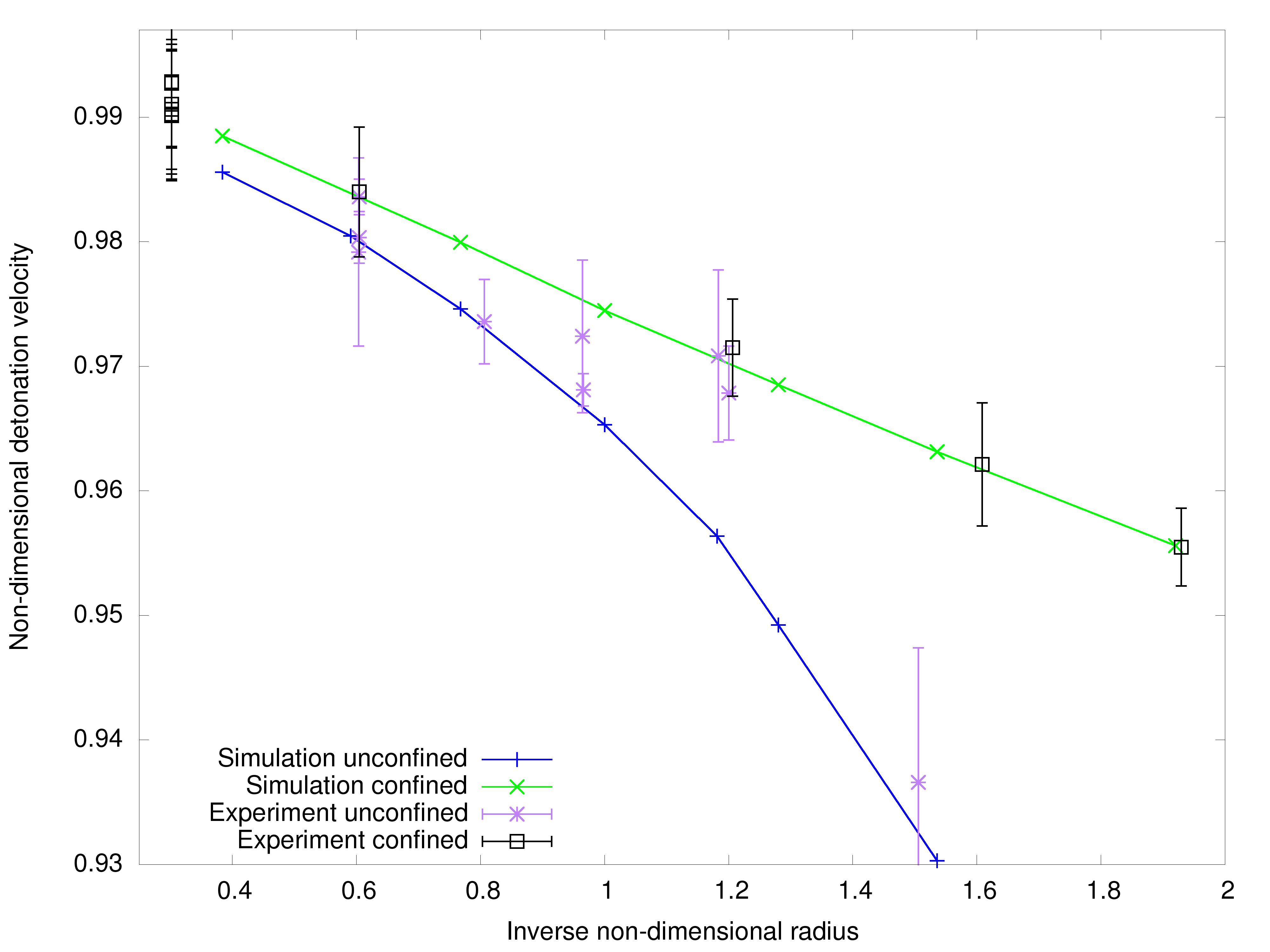}
 \caption{Comparison with the experiments of \citet{LX17Experiment} for the LX-17 rate stick test. The graph shows the non-dimensionalised steady state detonation velocity for various different stick radii in both the confined and unconfined tests. Good agreement is observed with experimental values, and a marked difference between confined and unconfined detonation velocity is seen.}
  \label{fig:LX17graph}
\end{figure}

\subsection{Rate Stick Tests}

The next test considered was a rate stick test, commonly used to assess reactive models \cite{Banks}, \cite{MiNi16}. This test features a cylinder of the condensed phase explosive LX-17 surrounded by either an elastoplastic copper container or air (the `confined' and `unconfined' cases respectively). The explosive is ignited by a booster and the reaction travels along the cylinder, eventually reaching a steady state detonation wave. The test is performed with different confining materials to asses the impact of the confiner on the steady state detonation velocity. As LX-17 is a slightly non-ideal explosive, a difference in the confined and unconfined steady state detonation velocities is observed experimentally by \citet{LX17Experiment}. This test is validated by performing a range of experiments with different rate stick radii, in both the confined and unconfined cases, and comparing to the results of \citet{LX17Experiment}.

The explosive's products and reactants are governed by the JWL equation-of-state previously outlined, with parameters given in Table \ref{tab:JWL}. The ignition and growth rate law was used to describe the explosive. More details of the use of this rate law for LX-17 are outlined in \citet{TarverLX17HockeyPuck}. This is a three-stage rate law, based on phenomenological experience of how detonations in condensed phase explosives evolve. The rate can be expressed as:
\begin{eqnarray}
\dot{\lambda} = I(1-F)^b\left(\frac{\rho}{\rho_0}-1-a\right)^xH(F_{ig}-F) \\
               + G_1(1-F)^cF^dp^yH(F_{G_1}-F) \nonumber \\
               + G_2(1-F)^eF^gp^zH(F-F_{G_2}) \nonumber
\end{eqnarray}
where $F = 1-\lambda$ is the reacted fraction and $H$ is the Heaviside function. Other parameters are material dependent constants, given in Table \ref{tab:IandG}.
A cylindrically symmetric domain was used for these tests, with a reflective boundary condition on the axis. A CFL number of 0.6 was used. The explosive cylinder had a radius varying from $r_0 = 4$ to $20$ mm, with a domain of $r = [0,2r_0]$, $z = [0,10r_0] $. AMR is required for this test, as the resolution of the detonation front is crucial in measuring accurate detonation velocities. A base resolution of 100 $\times$ 500 is used, with 2 refinement levels each with a refinement factor of 2. Criteria are needed to choose where the spatial grid should be refined in order to reduce error and preserve flow features. A difference-based approach is taken, where when the fractional change in a refinement variable between a cell and its neighbour is larger than a specified amount, those cells are tagged for refinement in the procedure outlined previously. Generally this flagging is performed on the variables $\phi_{(l)}, \phi_{(l)}\rho_{(l)}, u_i,$ and $p$.

The solid was described by the Romenskii equation-of-state with the parameters laid out in Section \ref{sec:1DTest}, using ideal plasticity with a yield stress of $\sigma_y = 0.4 \ \mbox{GPa}$. The air, when used, was described by an ideal gas with $\gamma = 1.4$.

The detonation was started by a booster region at the end of the cylinder with a pressure of $27 \times 10^9$ Pa and a thickness of $r_0$, with the other materials at atmospheric pressure. The explosive is initialised with a density of $\rho = \rho_0 = 1905 \text{ kg$\cdot$m}^{-3}$.

It is worth reiterating here that this test is performed with all materials having diffuse interfaces between them. There is no use of level sets, only the THINC algorithm to keep the appropriate interfaces sharp. This contrasts the mixed approach used by both \citet{SchochCoupled} and \citet{MiNi16}.

The results are non-dimensionalised following the procedure in \citet{IoannouLX17CornerTurning}. Here, quantities are non-dimensionalised using the CJ (Chapman-Jouguet) detonation velocity, $D_{\mbox{\scriptsize{CJ}}}$, for LX-17 and a reference time:
\begin{align}
D_{\mbox{\scriptsize{CJ}}}  &= 7.6799 \times 10^{3} \  \mbox{m $\cdot$ s}^{-1} \\
t_{\mbox{ref}} &= 1 \ \mu \mbox{s} \\
l_{\mbox{ref}} &= D_{\mbox{\scriptsize{CJ}}} \cdot t_{\mbox{ref}} = 7.6799 \ \mbox{mm}
\end{align}

Figure \ref{fig:LX17graph} shows the results of several confined and unconfined rate stick tests with comparison against experiments of \citet{LX17Experiment}. The non-dimensionalised detonation velocity is plotted for several non-dimensionalised inverse radii. The results agree with experiment, demonstrating the marked difference between the two cases. The results for the confined case agree particularly well. The unconfined case is more difficult to match, as experiments show that for radii larger than around 6.5 mm (corresponding to a non-dimensionalised inverse radius of around 1.2) a steady detonation is not formed, leading to the drastic drop in velocity. However, this threshold is correctly reflected  in the results of this method. Importantly, there is no change in the numerical methods required for either the confined or unconfined case.

Figure \ref{fig:ratestick} shows the $r_0 = 20$ mm confined and unconfined test for reference. The tests also show good qualitative agreement with previous numerical rate stick tests, such as \citet{IoannouLX17CornerTurning}, the strong confinement case of \citet{Banks} and \citet{MiNi16}.

\subsection{Explosive filled copper vessel}

\begin{figure*}
 \centering
 \includegraphics[width = 0.7\textwidth]{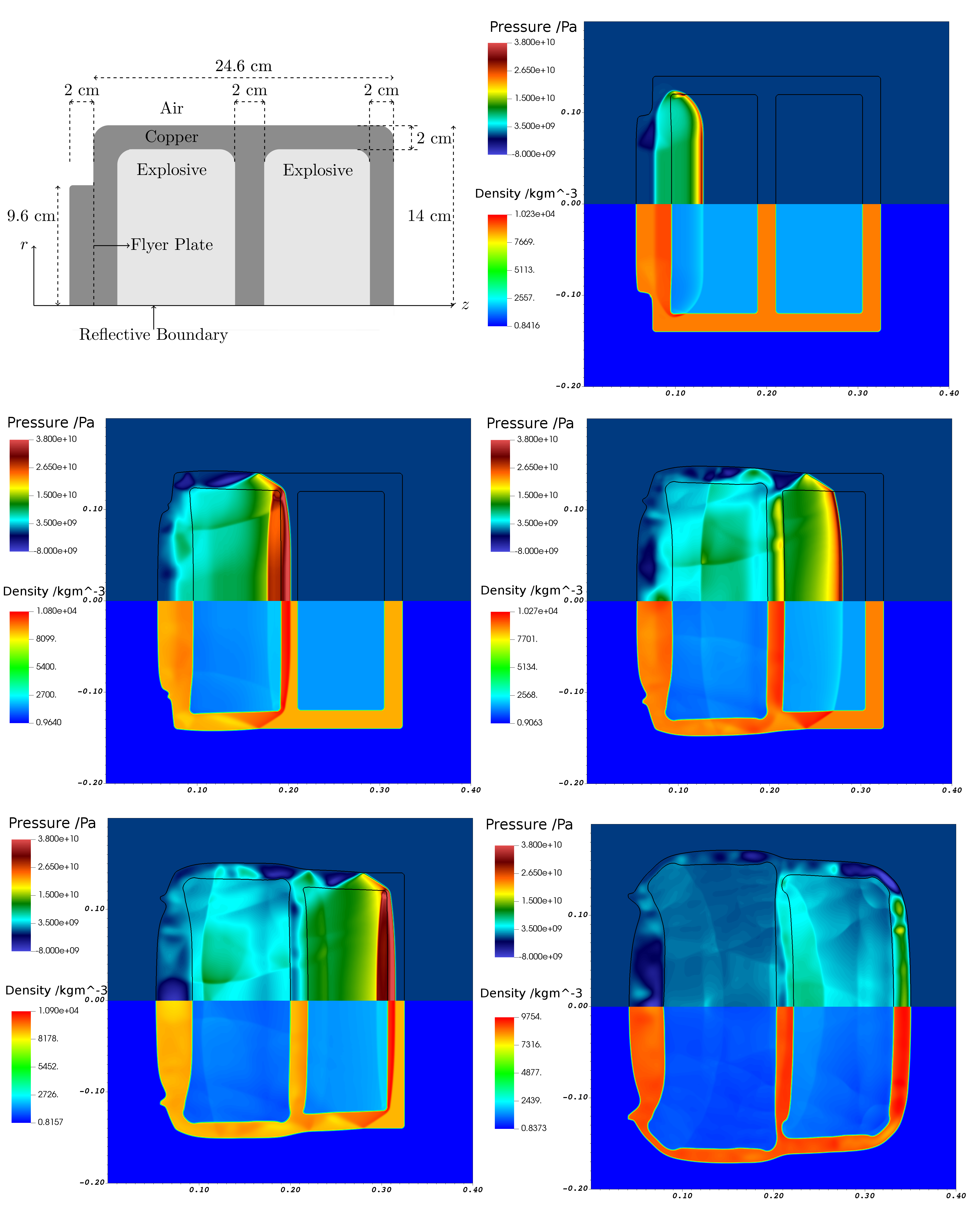}
 \caption{The \citet{SchochCoupled} shock-induced copper can test. The images are taken at t = 10, 19.5, 29, 33.5 and 54.5 $\mu$s, chosen to correspond to the results of \citet{SchochCoupled}. In each frame, the top shows the pressure and the bottom shows the density. This test demonstrates a full coupling of the elastoplastic and reactive mixture components of the model, showing good agreement with the sharp interface results of \citet{SchochCoupled}}
 \label{fig:FlyerPlate}
 \includegraphics[width = 0.77\textwidth]{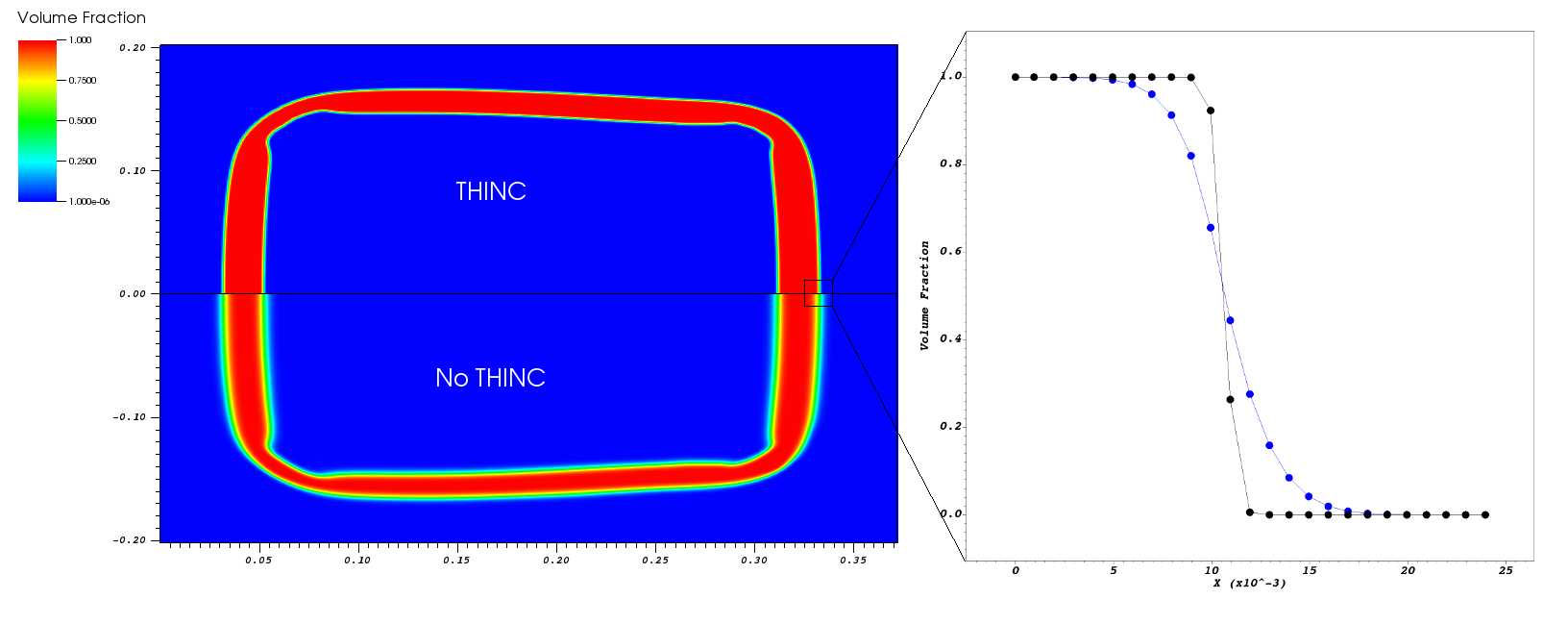}
  \caption{A comparison of THINC reconstruction and no reconstruction volume fraction profiles for the \citet{SchochCoupled} can test at 40 $\mu$s. The black points show the THINC profile and the blue points show the no reconstruction profile.}
 \label{fig:CanvolumeFraction}
\end{figure*}

\begin{figure}
 \centering
 \includegraphics[width = 0.5\textwidth]{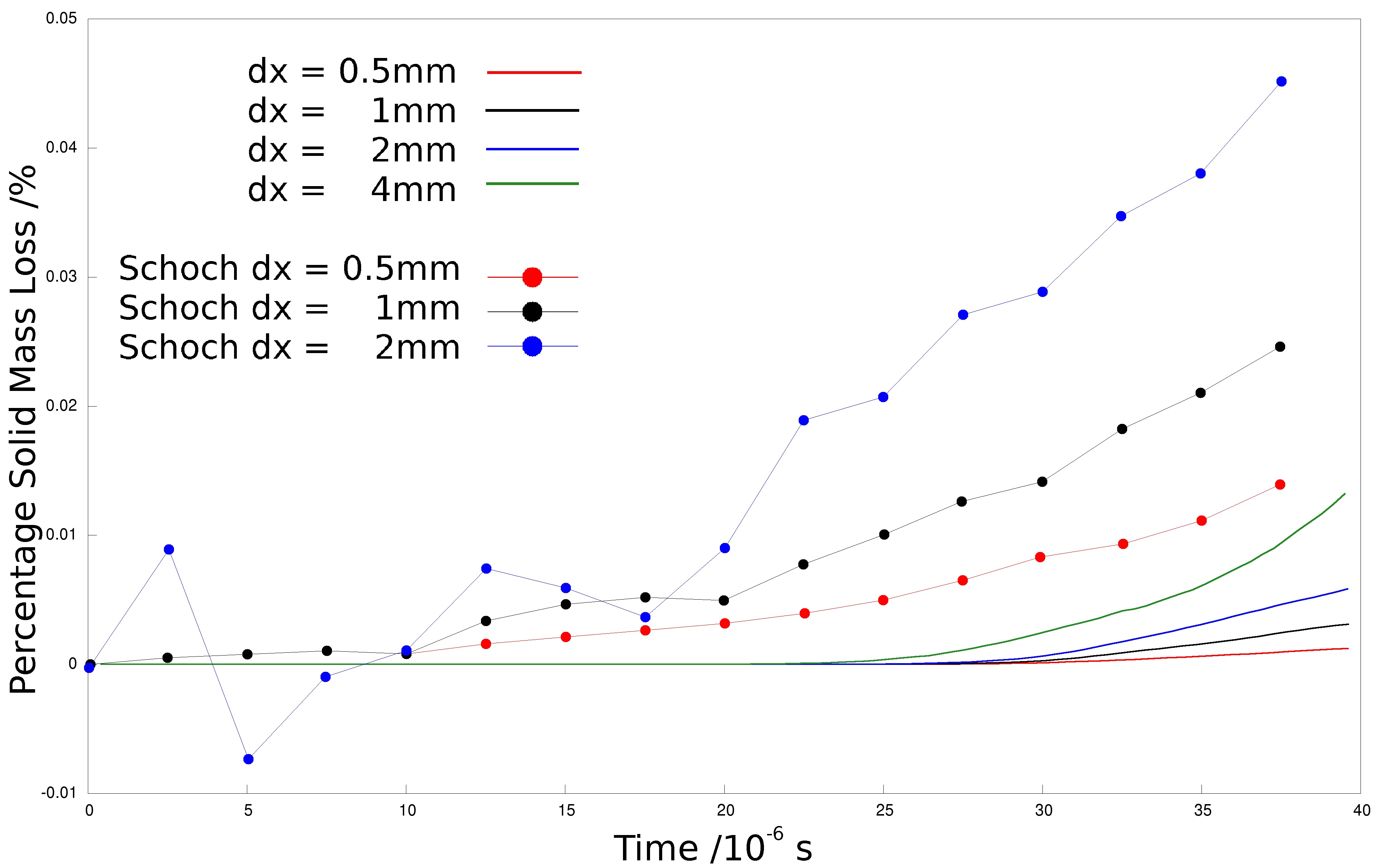}
 \caption{The percentage solid mass loss over the course of the \citet{SchochCoupled} explosive can test for different spatial resolutions. Data obtained by \citet{SchochCoupled} is plotted as points and this paper's data is plotted as lines. Corresponding colours represent the same resolution of test. In all cases, the mass loss incurred by this model is significantly lower than that of \citet{SchochCoupled}.}
 \label{fig:massChange}
\end{figure}

To more rigorously test the coupling of the explosive and elastoplastic solids and the ability of the THINC algorithm to maintain sharp interfaces over large deformations, the `explosive filled cooper vessel' tests from \citet{SchochCoupled} were considered. This test is commonly employed in detonation studies, and appears in a variety of forms, such as in \citet{Chinnayya}, \citet{MillerCollelaCoupled}, \citet{MiNi16} and \citet{MiNi18}.

Both the booster- and flyerplate-ignited versions of this test were considered. A cylindrically symmetric domain of $r = [0,20]$ cm, $z = [0,40] $ cm was used,  with a CFL number of 0.6 and a resolution of $N=400\times800$. The initial conditions for the flyerplate version of this test are set out in Figure \ref{fig:FlyerPlate}. The domain consists of a copper can filled with explosive, surrounded by air. In the test with the flyerplate, a partition is included in the centre of the can and the plate hits the can at a velocity of 400 m$\cdot$s$^{-1}$. For the booster ignited version, the partition is removed and the detonation is started by a 1 cm thick booster region with a pressure of 10$^9$ Pa.

The explosive here is the same idealised explosive as was used in the one-dimensional reactive test and the copper is again governed by the Romenskii equation-of-state. Following \citet{SchochCoupled}, this test uses ideal plasticity with a yield stress of $\sigma_Y = 0.07$ GPa.

Figure \ref{fig:FlyerPlate} shows how the pressure and density progressed through the flyerplate test. This method recovers the same behaviour as that found by \citet{SchochCoupled}; the shock is able to pass through the solid partition and reignite on the other side, a demonstration of the full multi-physics coupling of this model.

Figure \ref{fig:massChange} shows the percentage mass change for the solid phase in the booster ignited test, compared to the results obtained by \citet{SchochCoupled}. The method presented here out-performed the Ghost Fluid method at all resolutions tested. This is due to the intrinsic conservation errors in sharp interface methods that this scheme avoids. On top of this, Figure \ref{fig:CanvolumeFraction} shows how the THINC reconstruction helps to reduce numerical diffusion; evidently this is required when the copper confiner undergoes such large deformations.

\subsection{Hemispherical Indentation Tests}
\begin{figure}
 \centering
 \includegraphics[width = 0.4\textwidth]{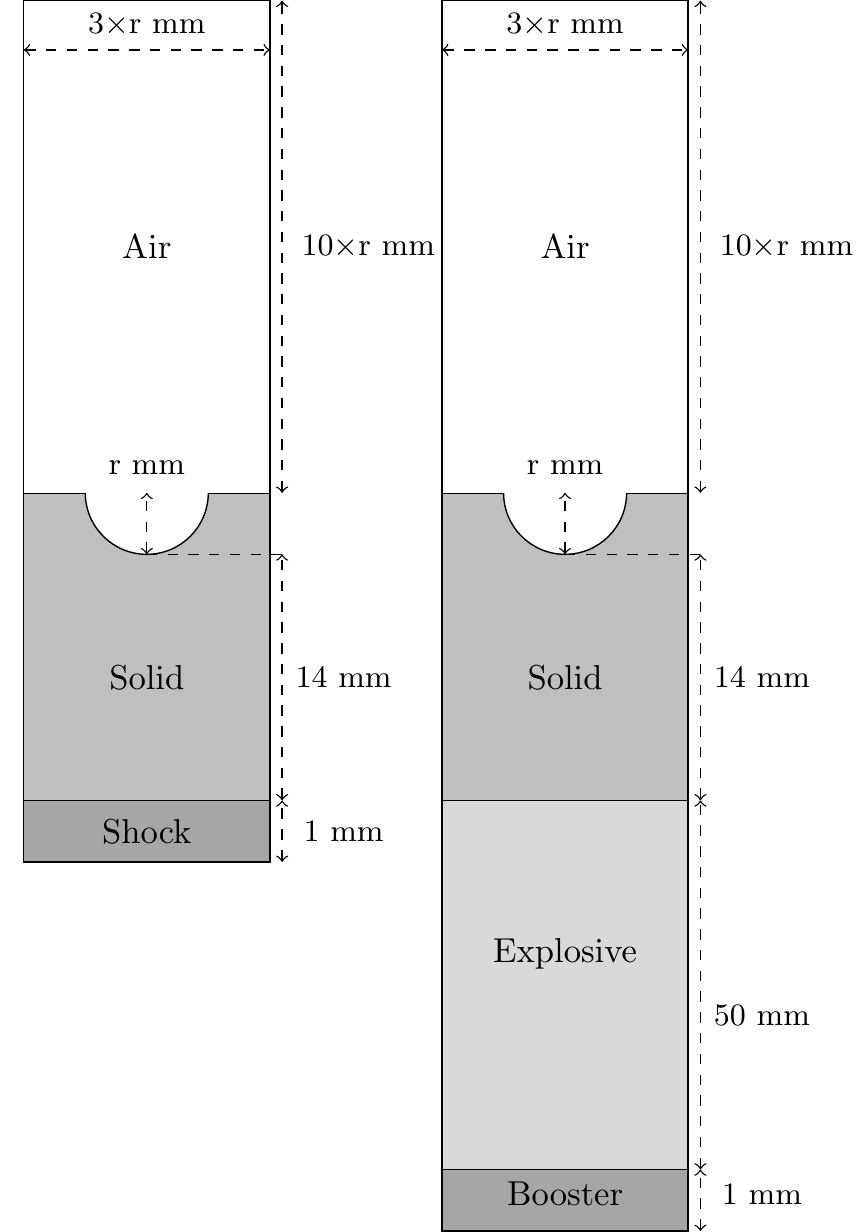}
 \caption{Simple shock (\textit{left}) and multiphase explosive (\textit{right}) initial conditions for the hemispherical indentation test with radius $r$.}
 \label{fig:GrooveInitialConditions}
\end{figure}

\begin{table}
\begin{center}
\begin{tabular}[t]{|c|c|c|c|c|c|}
  \hline
  $c_1$ / GPa & $c_2$ / GPa & $c_3$ & $n$ & $m$ & $T_{\text{melt}}$ / K \\
  \hline
  0.4 & 0.177 & 0.025 & 1.0  & 1.09 & 1358.0  \\
  \hline
\end{tabular}
\caption{The Johnson Cook plasticity parameters for copper, taken from \citet{UdaykumarGroove}.}~\\[0.2cm]
\label{tab:JCparameters}
\end{center}
\end{table}

To strenuously test all parts of the model, the final test considered is an explosively-initiated shock colliding with a hemispherical indentation in copper. This test has been performed experimentally by \citet{Mali_experiment_sphere} and numerically by \citet{UdaykumarGroove} and \citet{CooperGroove}, making it a suitable candidate for validation. Experimental research \cite{Mali_experiment_sphere,Mali_experiment_planar} has found that the problem has the benefit of being characterised by the radius of the indentation, if material parameters and initial conditions are kept constant, allowing for comparison between tests. When the shock hits the hemispherical indentation, the resulting converging wave causes a very high speed ($\approx $ Mach 10), high deformation, thin spike to jet out from the metal surface. Experimentally, the shock is generated by the detonation of an explosive impacting the metal. Previous numerical simulations have instead imposed shock conditions on the metal without the use of an explosive. This work includes both approaches. The shock conditions are considered first, where comparison can be made to other numerical experiments. Subsequently a multiphase explosive is used to drive the shock, demonstrating the capabilities of the model at hand, with the multi-physics approach enabling the full experimental set-up to be mimicked.

The copper in these tests used the Romenskii equation of state previously outlined, but now following the Johnson and Cook plasticity model \eqref{JCPlasticity} (with parameters outlined in Table \ref{tab:JCparameters}) to demonstrate the capability of the model to handle more complex, realistic plasticity models. The air was an ideal gas with $\gamma = 1.4$. When used, the explosive was the idealised condensed phase explosive used in the explosive-filled copper vessel tests, but with a reduced detonation energy of $Q=2.5$ MJ$\cdot$kg$^{-1}$. The tests employed a cylindrically symmetrical domain, with a reflective boundary condition on the axis centred on the indentation. A CFL of 0.6 was used for all the tests.

Two radii were considered: 4 mm and 15 mm. Both radii were initially tested with the shock initial conditions. The initial conditions for these tests are shown in Figure \ref{fig:GrooveInitialConditions}. The tests were run with a base resolution of 16 $\times$ 160, with 2 layers of AMR for the 4 mm test and 4 layers for the 15 mm test, giving a resolution of $1\times10^{-4}$ m in both cases. As mentioned by \citet{CooperGroove}, the experimental conditions for \citet{Mali_experiment_sphere} are very close the experiments performed by \citet{DeribasExplosiveModelling}. To this end, the shock conditions considered were taken from \citet{DeribasExplosiveModelling}, where the bottom of the domain was accelerated to 876 m$\cdot$s$^{-1}$. The thickness of the copper plate between the shock and the indentation was kept at 14 mm for both tests, meaning the only difference in either case was the radius of the indentation.

\begin{figure}
\centering
 \includegraphics[width = 0.5\textwidth]{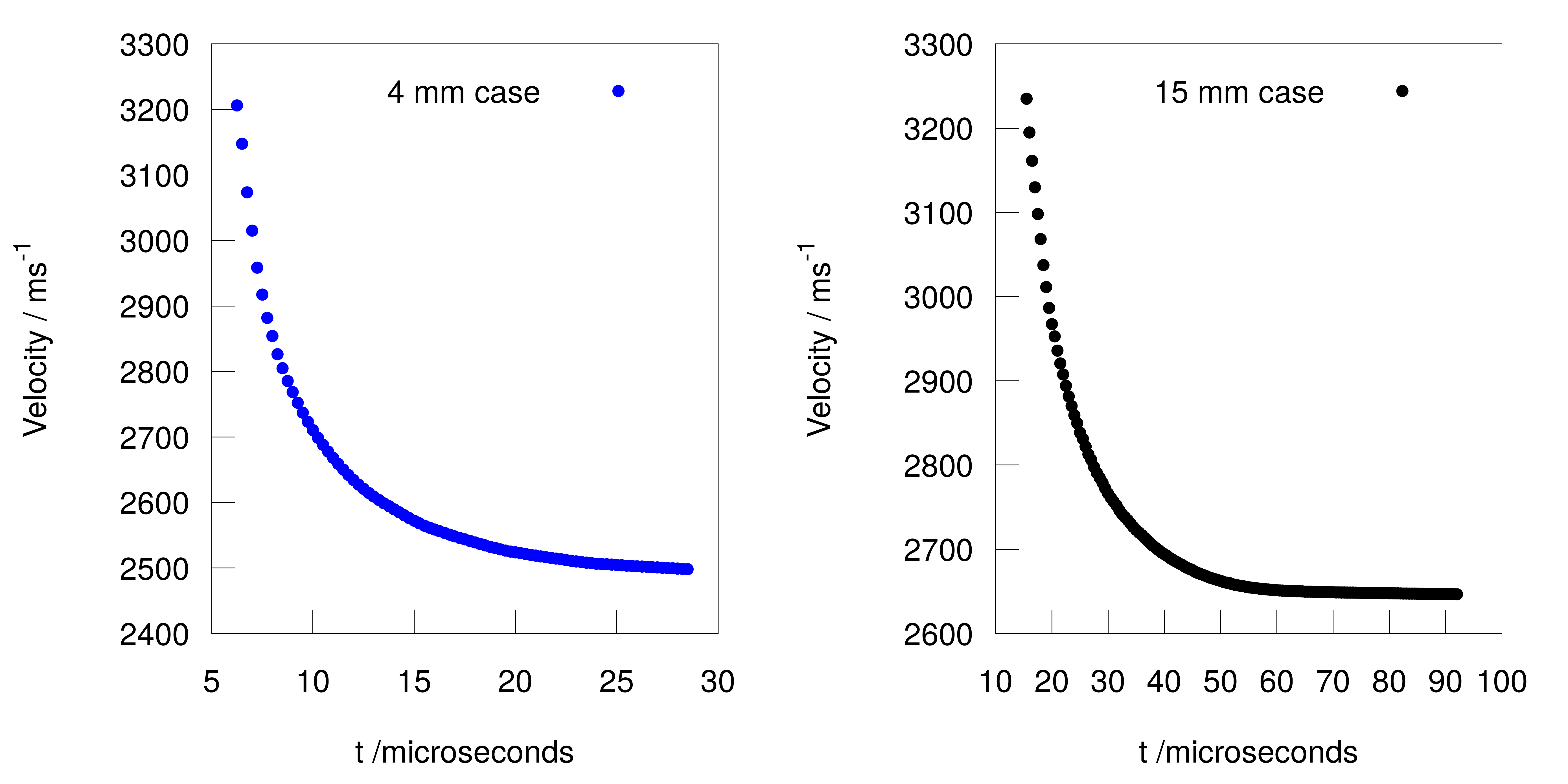}
 \caption{Jet velocity for the shock driven hemispherical indentation tests.}
 \label{fig:HG_data}
\label{fig:test}
\end{figure}

\begin{figure}
 \centering
 \includegraphics[width = 0.5\textwidth]{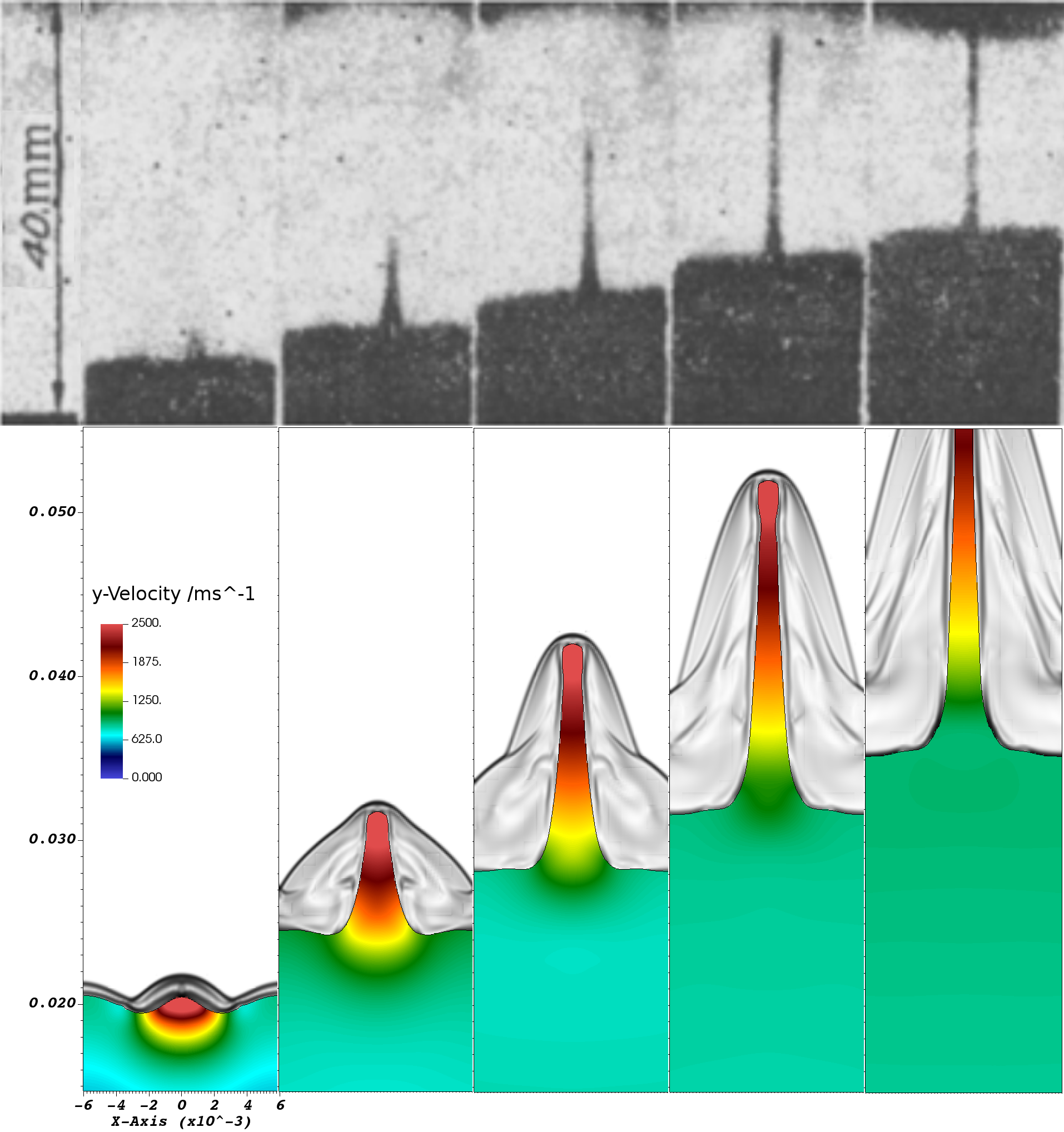}
 \caption{Experimental comparison with the work of \citet{Mali_experiment_sphere} for the 4 mm hemispherical indentation test. Images are taken every 4 $\mu$s. The images show the upward velocity of the jet, along with a numerical schlieren of the density in the air.}
 \label{fig:HG_Experimental_comparison}
\end{figure}

Table \ref{tab:HG_results} and Figure \ref{fig:HG_data} show how these numerical tests compare with previous results in terms of the radius and velocity of the jet produced. The test at hand differs from the other numerical tests mentioned here, as the diffuse interface model requires the metal to be surrounded by air, rather than vacuum. This leads to a more pronounced deceleration of the jet after leaving the surface, but also allows for the visualisation of the shock waves in the surrounding air. The 4 mm shock test is shown for reference, alongside the experimental images from \citet{Mali_experiment_sphere} in Figure \ref{fig:HG_Experimental_comparison}. It can be seen from Table \ref{tab:HG_results} and Figure \ref{fig:HG_data} that the results obtained with this model are consistent with experiment. The value for the velocity of the jet is taken as the steady state value which the results approach towards the end of the simulation. The radius is measured using the same approach as \citet{CooperGroove}, who take an average along the jet just before it would hit the target in the experiment. For this work that corresponds to when the jet reaches a distance from the initial surface of the metal of 10 $\times$ r. However, there is a large variation in the radius of the jet over the simulation, both along its length and over time, and the radius does not converge to a steady value. The value of the radius stated here should therefore be taken as a representative measure in order to compare to other numerical experiments.  

\begin{table}
\begin{center}
\begin{tabular}{|c|c|c|}
\hline
Test                                      	 & Jet Velocity/km$\cdot$s$^{-1}$ & Jet Radius/mm \\ \hline \hline
4mm Radius & & \\ \hline
Experimental \cite{Mali_experiment_sphere} 	& 2.5                      & 0.8 \\ 
This work  				   	& 2.49 $\pm$ 0.05          & 0.81 $\pm$ 0.02 \\
\citet{UdaykumarGroove}         		& -                        & - \\ 
\citet{CooperGroove}           			& 2.03 $\pm$ 0.14          & 0.75 \\ \hline
\hline
15mm Radius & &\\ \hline
Experimental \cite{Mali_experiment_sphere} 	& 2.7                      & 3.0\\ 
This work  				   	& 2.64 $\pm$ 0.05          & 2.93 $\pm$ 0.02 \\
\citet{UdaykumarGroove}         		& 2.75                     & 2.9 \\
\citet{CooperGroove}           			& 2.52 $\pm$ 0.12          & 2.4 \\
\hline
\end{tabular}
\end{center}
\caption{Results of the hemispherical indentation test from different sources for the 4 mm and 15 mm radius cases. Good agreement with previous simulation and experimental work is seen.}
\label{tab:HG_results}
\end{table}

\begin{figure*}
\centering
 \includegraphics[width = 0.9\textwidth]{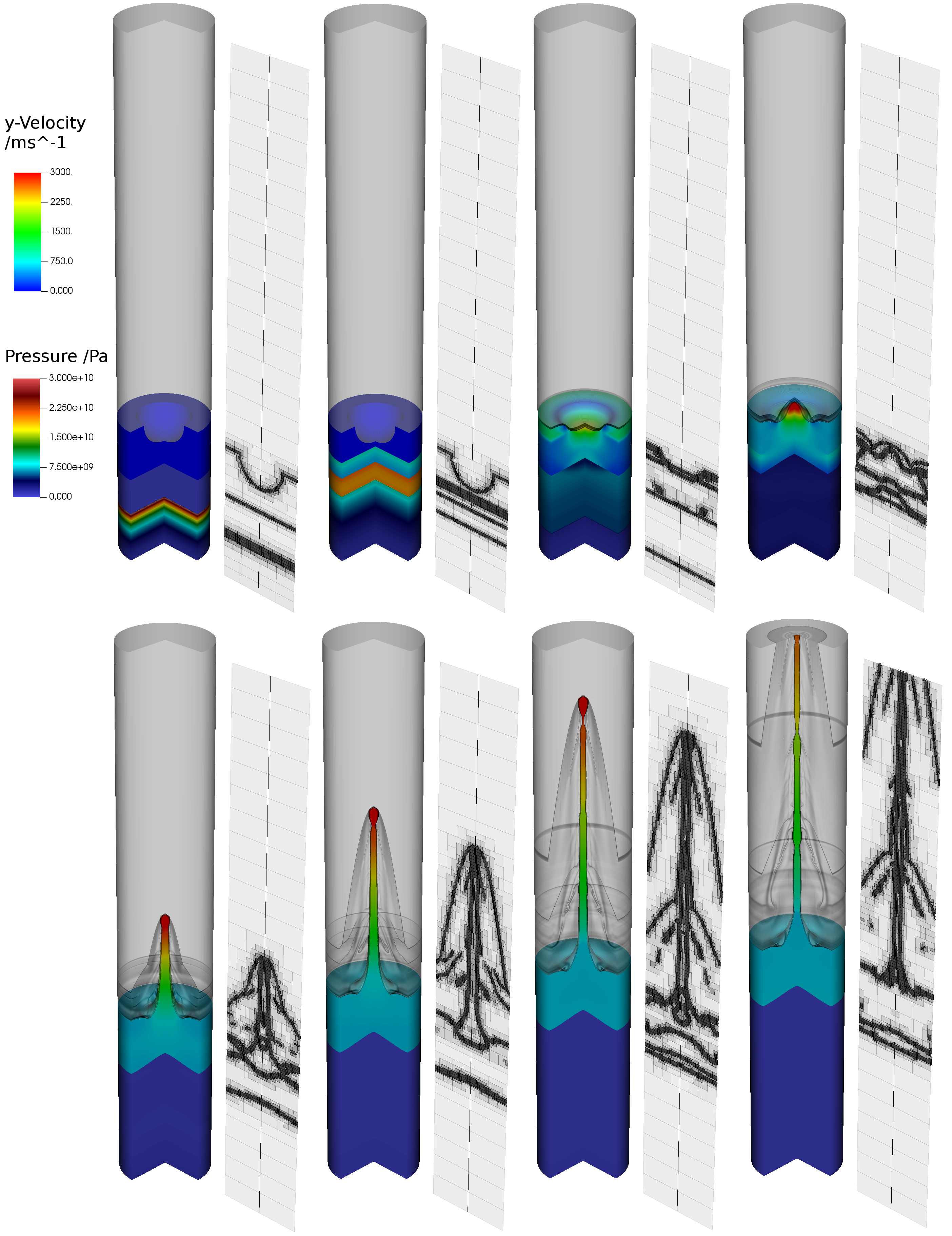}
 \caption{The 15 mm radius hemispherical indentation test. The images show the pressure in explosive, the upward velocity of the jet, and a numerical schlieren of the density in the surrounding air. The images are taken at 5, 10, 15, 20, 40, 60, 80 and 100 $\mu$s.}
 \label{fig:UdaykumarGroove_15mm}
\label{fig:test}
\end{figure*}

Having compared the shock driven case, the explosively driven case is detailed for a 15 mm radius indentation. The shock is now replaced by a multiphase explosive. The explosive is ignited with a booster region with pressure 1 GPa, and allowed to reach a steady state detonation wave before impacting the copper surface. The tests were run with a base resolution of 16 $\times$ 160, with 4 layers of AMR, using a CFL of 0.6. Figure \ref{fig:UdaykumarGroove_15mm} shows this test, depicting the pressure in explosive, the velocity in the solid, and a numerical schlieren of the shock waves in the surrounding air. The results are analogous to the shock driven case, but now also demonstrate the capability of the method to handle more realistic experimental conditions.

\section{Conclusion}
\label{sec:conclusion}

In this work, a multi-material, multi-phase, multi-physics diffuse interface scheme with sharpened material interfaces has been created and numerically validated against experiment and previous simulations. This was accomplished by combining diffuse interface elastoplastic methods, reactive fluid mixture methods, and THINC interface reconstruction techniques. The model is general, capable of handling an arbitrary number of materials that obey the broad class of Mie-Gr\"{u}neisen equations of state in three dimensions, one of which can be a reactive fluid mixture. This enables the direct, fully-coupled simulation of complex multi-material reactive-fluid problems, which have many valuable applications. Reactive-fluid simulations of this kind have previously been facilitated by either approximating elastoplastic solids as dense fluids solely through the equation of state, or by modelling the reactive mixtures as single phases with a programmed-burn style reaction rate. Both of these techniques omit physics that can be included by this model. Moreover, this combined system solves the same set of equations over the entire domain, with all materials being defined globally, without the need for a dividing level set. This avoids both having to use unphysical co-simulation techniques, or non-conservative sharp interface methods. This method is fully coupled and more conservative than sharp interface methods developed in recent times, and this model therefore presents a viable alternative method for complex multi-physics. It should also be added that the numerical methods presented here can easily be extended to a wide variety of other update- or flux-methods, or to encompass additional physics through the inclusion of additional equations to be solved in the same fashion.

\section*{Acknowledgements}

This work was funded by AWE PLC. Additionally, Tim Wallis was funded by a grant from the UK Engineering and Physical Sciences Research Council (EPSRC) EP/L015552/1 for the Centre for Doctoral Training (CDT) in Computational Methods for Materials Science.

\bibliography{mybibfile_edit}

\appendix

\section{THINC Reconstruction Algorithm}
\label{app:THINC}

The THINC algorithm models a thermodynamic quantity $q$ in a cell $i$ near an interface as obeying a $\tanh$ profile, given by:
\begin{align}
 q_i(x)^{\mbox{\scriptsize{THINC}}}  = q_{\mbox{\scriptsize{min}}}+\frac{q_{\mbox{\scriptsize{max}}}}{2}\left( 1+\theta \tanh\left(\beta\left(\frac{x-x_{i-\frac{1}{2}}}{x_{i+\frac{1}{2}}-x_{i-\frac{1}{2}}} - x_0 \right)\right)\right) \ .
\end{align}
Here $q_{\mbox{\scriptsize{min}}} = \mbox{min}(q_{i-1},q_{i+1})$, $q_{\mbox{\scriptsize{max}}} = \mbox{max}(q_{i-1},q_{i+1})- q_{\mbox{\scriptsize{min}}}$, $\theta = \mbox{sgn}( q_{i+1}- q_{i-1} )$ and $\beta$ is a parameter that controls the thickness of the interface. $ x_0 $ is the unknown interface location as a fraction of the cell width, such that when $x_0 = 0$ the interface lies at $x_{i-\frac{1}{2}}$, and when $x_0=1$ it lies at $x_{i+\frac{1}{2}}$. This interface location is found by solving 
\begin{align}
 q_i = \frac{1}{\Delta x} \int_{x_{i-\frac{1}{2}}}^{x_{i+\frac{1}{2}}} q_i(x)^{\mbox{\scriptsize{THINC}}} \dd{x} \ . \label{eq:THINCintegral}
 \end{align}
Using this form also ensures conservation of the variable. The cell-edge values are computed as
\begin{subequations}
\begin{align}
 q_i^L& = q_{\mbox{\scriptsize{min}}}+\frac{q_{\mbox{\scriptsize{max}}}}{2}\left( 1+\theta A\right) \\
 q_i^R& = q_{\mbox{\scriptsize{min}}}+\frac{q_{\mbox{\scriptsize{max}}}}{2}\left( 1+\theta \frac{\tanh(\beta)+A}{1+A\tanh(\beta)}\right) \ ,
\end{align}
\end{subequations}
where
\begin{align}
 A &= \frac{\frac{B}{\cosh(\beta)}-1}{\tanh(\beta)}, \\
 B &= \exp(\theta \beta(2C-1) ), \\
 C &= \frac{q_i-q_{\mbox{\scriptsize{min}}} + \epsilon}{q_{\mbox{\scriptsize{max}}}+\epsilon} \ , 
\end{align}
are constants that arise in the integration of Equation \ref{eq:THINCintegral}. A small positive quantity, $\epsilon$, is introduced to prevent division by zero. A value of $\epsilon = 10^{-20}$ is used in this work.

The decision of whether to apply this reconstruction is based on criteria for whether the cell in question is a \textit{mixed cell}, defined as satisfying:
\begin{align}
 \delta < C < 1-\delta, \ \ (q_{i+1}-q_i)(q_i-q_{i-1}) > 0 \ ,
\end{align}
where $\delta$ is a small positive value. This work uses $\delta = 10^{-5}$. 

Following this initial reconstruction along the lines of \citet{XiaoTHINC}, the BVD (Boundary Variation Diminishing) algorithm of \citet{BVDTHINC} is then applied in conjunction with a MUSCL extrapolation scheme. The premise of this scheme involves minimising the variation between cells, hence reducing the dissipation at interfaces, by comparing the THINC reconstruction with the MUSCL extrapolation and taking whichever gives a lower boundary variation. This is achieved by comparing the \textit{total boundary variation} (TBV) of each scheme, given by:
\begin{eqnarray}
 \mbox{TBV}^{\mbox{\scriptsize{P}}} &= \min ( \ &\lvert q_{i-1,R}^{\mbox{\tiny{MUSCL}}}-q_{i,L}^{\mbox{\tiny{P}}}\rvert+\lvert q_{i,R}^{\mbox{\tiny{P}}}-q_{i+1,L}^{\mbox{\tiny{MUSCL}}}\rvert ,  \nonumber \\
 &&\lvert q_{i-1,R}^{\mbox{\tiny{THINC}}}-q_{i,L}^{\mbox{\tiny{P}}}\rvert+\lvert q_{i,R}^{\mbox{\tiny{P}}}-q_{i+1,L}^{\mbox{\tiny{THINC}}}\rvert,\nonumber\\
 && \lvert q_{i-1,R}^{\mbox{\tiny{MUSCL}}}-q_{i,L}^{\mbox{\tiny{P}}}\rvert+ \lvert q_{i,R}^{\mbox{\tiny{P}}}-q_{i+1,L}^{\mbox{\tiny{THINC}}}\rvert , \nonumber \\
 &&\lvert q_{i-1,R}^{\mbox{\tiny{THINC}}}-q_{i,L}^{\mbox{\tiny{P}}}\rvert+ \lvert q_{i,R}^{\mbox{\tiny{P}}}-q_{i+1,L}^{\mbox{\tiny{MUSCL}}}\rvert ) \ ,
\end{eqnarray}
where $\mbox{P}$ stands for either MUSCL or THINC. Therefore the final criterion for whether the THINC reconstruction is applied to a cell is given by:
\begin{align}
 q_i^{\mbox{\scriptsize{BVD}}} = \left\lbrace \mqty{q_i^{\mbox{\scriptsize{THINC}}} & \mbox{ if } \mbox{TBV}^{\mbox{\tiny{THINC}}} < \mbox{TBV}^{\mbox{\tiny{MUSCL}}} \\ q_i^{\mbox{\scriptsize{MUSCL}}} & \mbox{ otherwise} }  \right. \ .
\end{align}

Multidimensionality is accounted for following the procedure used by \citet{XiaoTHINC}, which takes
\begin{align}
 \beta = \beta_0  \lvert n_d \rvert +0.01 \ ,
\end{align}
where $\beta_0 = 2.5$ is a constant and $\lvert n_d \rvert$ is the magnitude of the component of the interface normal in a given direction. This normal vector is calculated using Youngs' method \cite{YoungsNormal}.

\section{Root Finding procedure}

\label{app:RootFinding}
The root finding procedure is as follows. The mixture rule for the total internal energy \eqref{Energy_mixture_rule} and the mixture rule for internal energies in a physical mixture \eqref{eq:E_physicalmixturerule} are combined:
\begin{align*}
\rho \mathscr{E} &= \left(\sum_{x} \phi_{x}\rho_{x}\mathscr{E}_{x}\right) + \phi_{y}\rho_{y}\left(\lambda_{y} \mathscr{E}_{\alpha}+(1-\lambda_{y})\mathscr{E}_{\beta}\right) \ .
\end{align*}
Here $x$ refers to all non-mixture materials (which do not require root finding) and $y$ represents the mixture material. The goal is then to find the values of the pressure $p$ and the partial mixture densities, $\rho_{\alpha}$ and $\rho_{\beta}$, for which this equation holds.
Substituting the form of the equation-of-state \eqref{eq:MieGruneisenEOS}, along with the assumption that pressure is constant between materials, gives:
\begin{align*}
\rho \mathscr{E} =& \left(\sum_{x}\phi_{x}\rho_{x}\left(\frac{p-p_{\Subref,x}(\rho_{x})}{\rho_{x}\Gamma_x} +  \mathscr{E}_{\Subref,x}(\rho_{x}) \right)\right) + \cdots \\
&\phi_{y}\rho_{y}\lambda_{y}\left(\frac{p-p_{\Subref,\alpha}(\rho_{\alpha})}{\rho_{\alpha}\Gamma_\alpha} +  \mathscr{E}_{\Subref,\alpha}(\rho_{\alpha}) \right) + \cdots \\
&\phi_{y}\rho_{y}(1-\lambda_{y})\left(\frac{p-p_{\Subref,\beta}(\rho_{\beta})}{\rho_{\beta}\Gamma_\beta} +  \mathscr{E}_{\Subref,\beta}(\rho_{\beta}) \right) \ .
\end{align*}
Rearranging this equation for the pressure, $p$, gives:
\begin{align}
p = &\frac{1}{\left(\sum_{x} \frac{\phi_x}{\Gamma_x}\right) + \frac{\phi_y\rho_y\lambda_y}{\rho_{\alpha}\Gamma_{\alpha}} + \frac{\phi_y\rho_y(1-\lambda_y)}{\rho_{\beta}\Gamma_{\beta}}} \times \cdots \nonumber \\
&\Bigg[ \rho \mathscr{E} + \left(\sum_{x} \phi_x\rho_{x}\left(\frac{p_{\Subref,x}(\rho_x)}{\rho_{x}\Gamma_x} - \mathscr{E}_{\Subref,x}(\rho_x)\right)\right) + \cdots \nonumber \\
& \phi_y\rho_y\lambda_y \left( \frac{p_{\Subref,\alpha}(\rho_{\alpha})}{\rho_{\alpha}\Gamma_{\alpha}} - \mathscr{E}_{\Subref,\alpha}(\rho_{\alpha})\right)+ \cdots \nonumber \\
& \phi_y\rho_y(1-\lambda_y) \left( \frac{p_{\Subref,\beta}(\rho_{\beta})}{\rho_{\beta}\Gamma_{\beta}} - \mathscr{E}_{\Subref,\beta}(\rho_{\beta})\right)\Bigg] \ . \label{eq:pressureFunction}
\end{align}
From here, equation \eqref{eq:rho_physicalmixturerule} is manipulated to give an expression for $\rho_{\beta}$ in terms of $\rho_{\alpha}$ and $\rho_{y}$:
\begin{align}
 \rho_{\beta} = \frac{(1-\lambda_y)}{\frac{1}{\rho_y} - \frac{\lambda_{(y)}}{\rho_{\alpha}}} \ .
\end{align}
Finally, temperature is assumed to be constant between species in a physical mixture. Temperature is given by:
\begin{align}
 T_{(l)} = \frac{p - p_{\Subref, (l)}}{\rho_{(l)}\Gamma_{(l)}C^V_{(l)}} \ ,
\end{align}
where $C_{V}$ is the specific heat at constant volume for the material. Here $l$ can equally refer to an inert material or a reactant or product of a mixture. This assumption can be stated as:
\begin{align}
 T_{\alpha} &= T_{\beta} \\
 \frac{p - p_{\Subref, \alpha}}{\rho_{\alpha}\Gamma_{\alpha} C^V_{\alpha}} &= \frac{p - p_{\Subref, \beta}}{\rho_{\beta}\Gamma_{\beta} C^V_{\beta}} \label{eq:temperatureMix}
\end{align}
Equation \eqref{eq:temperatureMix} can now be written purely in terms of $\rho_{\alpha}$ and other known variables. Now the root finding procedure may begin. An initial guess for the density $\rho_{\alpha}$ is made and $\rho_{\beta}$ is calculated. To find the pressure, the method diverges depending on whether the conservative or the primitive variables are currently known. If only the conservative variables are known, the pressure is calculated with Equation \ref{eq:pressureFunction}. If the primitive variables are known, then the pressure can be used directly. The two mixture temperatures are calculated and compared using Equation \ref{eq:temperatureMix}. When the two temperatures are equal, the densities are correct. 
A robust bisection root-finding technique is used for this task, but a more efficient method could be implemented if desired. 
\section{Symbols}
\begin{table}
\begin{center}
\begin{tabular}{|l|l|}
\hline
Symbol & Meaning \\ \hline
$\phi$ & Volume fraction \\ \hline
$\rho$ & Total density \\ \hline
$\rho_{(l)} $ & Phasic density \\ \hline
$\vb{u}, u_k $ & Velocity \\ \hline
$\boldsymbol\sigma, \sigma_{ij} $ & Cauchy stress tensor \\ \hline
$E$ & Specific total energy \\ \hline
$\mathscr{E}$ & Specific internal energy \\ \hline
$p$ & Pressure \\ \hline
$T$ & Temperature \\ \hline
$\theta_D$ & Non-dimensional Debye Temperature \\ \hline
$c$ & Speed of sound \\ \hline
$Y$ & Mass fraction \\ \hline
$C^{V}$ & Specific heat capacity \\ \hline
\hline Mixtures & ~ \\ \hline
$\rho_{\alpha},\rho_{\beta}$ & Mixture densities \\ \hline
$\lambda$ & Reaction progress variable \\ \hline
$Q$ & Detonation energy \\ \hline
\hline Elastoplastic solids & ~ \\ \hline
$\Vbar$ & Left unimodular elastic stretch tensor \\ \hline
$\mathbf{H}^e$ & Hencky strain tensor \\ \hline
$\cal{J}$ & Second invariant of the shear strain \\ \hline
$G$ &Shear modulus \\ \hline
$\varepsilon_p$ & Plastic strain \\ \hline
$\chi$ & Plastic strain rate \\ \hline
$\Phi$ & Representation of plastic effects \\ \hline
$\sigma_Y$ & Plastic yield stress \\ \hline
$c_1,c_2,c_3,n,m,T_0,T_{\mbox{melt}}$ & Johnson Cook plasticity parameters \\ \hline
\hline Equations of State & ~ \\ \hline
$\gamma$ & Adiabatic index \\ \hline
$\Gamma$ & Mie-Gr\"{u}neisen coefficient \\ \hline
$\rho_0$ & Reference density \\ \hline
$p_{\Subref}, \mathscr{E}_{\Subref}$ & Mie-Gr\"{u}neisen reference functions \\ \hline
$p_{\infty}, e_{\infty}$ & Stiffened gas parameters \\ \hline
${\cal A},{\cal B},{\cal R}_1,{\cal R}_2$ & JWL fitting parameters \\ \hline
$K_0, G_0$ & Reference bulk and shear moduli\\ \hline
$\bar{\alpha},\bar{\beta}$ & Romenskii equation of state exponents \\ \hline
\end{tabular}
\end{center}
\end{table}

\end{document}